\documentclass[onecollarge,natbib]{svjour2}
\bibpunct{[}{]}{;}{n}{}{,} 
\smartqed  
\usepackage{graphicx}
\newcommand{\Q}{{\cal Q}}
\newcommand{\xB}{x_{\rm B}}
\newcommand{\GeV}{\rm GeV}
\journalname{Few-Body Syst}
\usepackage{amsmath}
\usepackage{amssymb}
\usepackage{xcolor}
\usepackage{tensor}
\usepackage{braket}
\usepackage{units}
\usepackage[hypertex,colorlinks=true,linkcolor=red,citecolor=blue]{hyperref}

\begin{document}

\title{Generalized Parton Distributions\thanks{Based on talks given at conferences/workshops:  Venturing off the lightcone - local versus global features (Light Cone 2013), May 20-24, 2013, Skiathos, Greece;
Exploring QCD with Next Generation Facilities (QCD Frontier 2013), Thomas Jefferson National Accelerator Facility,  October 21-22, 2013, Newport News, US;
Deeply Virtual Compton Scattering: From Observables to GPDs, Ruhr-Universit\"at Bochum, February 10-12, 2014, Bochum, Germany;
Studies of 3D Structure of Nucleon, Institute for Nuclear Theory, February 24-28, 2014, Seattle, US.
I am deeply indebted to the organizers of these conferences for support.}
}
\subtitle{-- visions, basics,  and realities --}


\author{D.~M\"uller}


\institute{D.~M\"uller \at
              Institute of Theoretical Physics II, Ruhr-University Bochum, D-44780 Bochum\\
              Tel.: +49 234 32-23728\\
              Fax: +49 234 32-14697\\
              \email{dieter.mueller@tp2.rub.de}           
}

\date{
}

\maketitle

\begin{abstract}
An introductory to generalized parton distributions is given which emphasizes their spectral property and its uses as well as the
equivalence of various GPD representations.
Furthermore, the status of the theory and  phenomenology of hard exclusive processes is shortly reviewed.

\keywords{Generalized Parton Distributions, Deeply Virtual Compton Scattering, Deeply Virtual Meson Production, Light-Front Wave Functions}
\end{abstract}

\section{Introduction}
\label{intro}

Generalized parton distributions (GPDs) were conceptually introduced in connection with the partonic description of
deeply virtual Compton scattering (DVCS) \cite{Mueller:1998fv,Radyushkin:1996nd,Ji:1996nm} and
deeply virtual meson production (DVMP) \cite{Radyushkin:1996ru,Collins:1996fb}.
In leading power w.r.t.~the inverse photon virtuality these processes factorize in a perturbatively calculable
hard-scattering part and universal, i.e., process-independent, however, conventionally defined GPDs.
Factorization theorems were derived for these processes to the leading twist-two level,
i.e., for transverse polarized photons in DVCS  \cite{Collins:1996fb} and longitudinal polarized photons in DVMP
\cite{Collins:1998be}.

The broad interest on GPDs arises from the fact that they encode
non-perturbative dynamics of the constituents, i.e., partons  in
hadrons or even nuclei, on the amplitude level. In fact they might
be interpreted as an overlap of light-front wave functions
\cite{Diehl:1998kh,Diehl:2000xz,Brodsky:2000xy}. Adding to the
extensive studies of form factors and parton distribution functions (PDFs), GPDs offer
a complementary insight into the nucleon and nuclei. Comprehensive reviews about GPDs, their interpretation, and the
phenomenology are given in Refs.~\cite{Diehl:2003ny,Belitsky:2005qn}. We would
like here to point out only the most two important goals of GPD
phenomenology. For vanishing skewness, i.e., $\eta=0$, their
Fourier transforms to the impact parameter space possess in the
infinite momentum frame a probabilistic interpretation
\cite{Burkardt:2000za,Burkardt:2002hr}. Namely, they give the probability to find a
parton with a given momentum fraction $x$ and transversal distance
from the nucleon center. We emphasize that parton densities,
intensively studied in inclusive processes, can only provide the
parton distribution with respect to the longitudinal momentum
fraction. We consider  a three-dimensional resolution of the
nucleon (and other hadrons) or even nuclei as a feasible task that
can be already undertaken at present in a model dependent manner.  Another novel application
arises from the fact that their second Mellin moment with respect
to $x$ gives the fraction of angular momentum carried by the
particular parton flavor \cite{Ji:1996ek}. Consequently, one of the
main goals of the GPD phenomenology  is the resolution of the spin
decomposition of the proton in partonic degrees of freedom. So far
the spin content of the nucleon, in particular of the sea quark and gluon contributions,
is not well understood.

Triggered by these promises,
the experimental efforts in measuring DVCS and DVMP processes during the
last few years led to an ever increasing amount and precision of data
for various observables. In particular, data have been taken
by the HERA collaborations in the fixed target experiment HERMES,
(DVCS: \cite{Airapetian:2001yk,Airapetian:2006zr,Airapetian:2008aa,Airapetian:2009aa,Airapetian:2010ab,Airapetian:2011uq,Airapetian:2012mq,Airapetian:2012pg},
DVMP: \cite{Airapetian:2000ni,Airapetian:2009ad,Airapetian:2007aa,Airapetian:2009ac} )
and the collider experiments H1 and ZEUS, e.g.,
(DVCS: H1 \cite{Adloff:2001cn,Aktas:2005ty,Aaron:2007ab,Aaron:2009ac}, ZEUS \cite{Chekanov:2003ya, Chekanov:2008vy}, DVMP: \cite{Aid:1996ee,Adloff:1997jd,Aaron:2009xp,Breitweg:1998nh,Breitweg:2000mu,Chekanov:2005cqa,Chekanov:2007zr}) at the
Deutsches Elektronen Synchrotron (DESY).  Precession measurements are
performed within dedicated experiments from the HALL A and CLAS
collaborations at the Thomas Jefferson National Laboratory (JLAB) (CLAS \cite{Stepanyan:2001sm,Chen:2006na,Girod:2007aa,Gavalian:2008aa} and Hall A \cite{Munoz_Camacho:2006hx,Mazouz:2007aa},
DVMP: \cite{Hadjidakis:2004zm,Morand:2005ex,Morrow:2008ek,Santoro:2008ai,DeMasi:2007id,Bedlinskiy:2012be,Blok:2008jy})
and new data sets are expected to be analyzed and released soon. Further measurements are planed after the $12\, \GeV$ upgrade.
Moreover, experimental measurements with a muon beam are planned to be
performed by the COMPASS II collaboration at the Organisation
Europ\'eenne pour la Recherche Nucl\'eaire (CERN), Switzerland.

The perturbative factorization approach has been confronted
with simple GPD ans\"atze that are adjusted to phenomenological PDF parametrizations and form factor data. The
first measurements of photon electroproduction were easily
understood at leading order (LO) in terms of oversimplified GPD models,
where the momentum fraction and momentum transfer square
dependence are factorized in terms of PDFs and form factors, e.g., see \cite{Goeke:2001tz,Belitsky:2001ns}.
In particular, the sizable first DVCS beam spin asymmetry measurements \cite{Airapetian:2001yk,Stepanyan:2001sm} were reproduced.
However, it became also clear that the
cross section normalization of various hard exclusive processes at
large energies, predicted to LO, overshoot the
measured ones of H1 and ZEUS collider experiments and that even for DVCS the quantitative GPD
description of various observables is not satisfactory \cite{Kumericki:2008di,Guidal:2013rya}. To overcome these phenomenological issues, a global GPD
fitting framework is under development, which is based on flexible GPD models \cite{Kumericki:2007sa,Mueller:2013caa}.
Presently, a good description of unpolarized DVCS world data set, including DVMP data at small $\xB$ is reached \cite{Kumericki:2009uq,Kumericki:2011zc,Lautenschlager:2013uya} and
a reasonable description of polarized DVCS  data could be achieved, too \cite{Kumericki:2013br}. Note that such a good description
of DVCS observables has  not been succeeded with popular GPD models or even with inconsistent ones.

The outline of the presentation is as follows. In Sec.~\ref{sec:1} we define twist-two GPDs and discuss  their representations
as well as the uses of the spectral representation, which we exemplify by means of a photon GPD.
In Sec.~\ref{sec:2} we recall the status of the theory and GPD modeling as used in phenomenology.
In Sec.~ \ref{sec:4} we shortly describe the phenomenological situation. Finally, we summarize.

\section{Definition, properties, and representations of generalized parton distributions}
\label{sec:1}
GPDs  are defined as  hadronic matrix elements of renormalized light-ray operators which possess certain quantum numbers.
They are characterized by their (geometrical) twist, given as dimension minus spin. The leading twist-two quark and gluon
GPDs contain two parton fields that live on a light ray, given by a light-like vector $n^\alpha$ with $n^2=0$, and are separated by
the distance $2\kappa$. The following definitions \cite{Mankiewicz:1997bk,Belitsky:1999hf}, agreeing with those in \cite{Diehl:2003ny,Belitsky:2005qn}, are usually
adopted%
\footnote{Besides the standard QCD definitions we use $P=p_1+p_2$, $\Delta=p_2-p_1$, $\epsilon^{\perp}_{\mu\nu} = \epsilon_{\mu\nu\alpha\beta}\, n^{\ast\alpha} n^\beta$
with $\epsilon^{0123}=1$ and $n^{\ast}\cdot n =1$. Furthermore, we do not display the path ordered gauge link that connects the field operators along the
light cone, which is absent in the light-cone gauge $n\cdot A =0$.   }
\begin{subequations}
\label{eq:defGPDV}
\begin{eqnarray}
 q(x, \eta, t,\mu^2)& =&   \int \frac{d \kappa}{2 \pi}\: e^{i x (P\cdot n)\kappa}
 \langle s_2,p_2|\bar{q}(-\kappa n)\, n\cdot\gamma\, q(\kappa n)|p_1,s_1\rangle_{(\mu^2)},
\label{eq:defquarkGPDV} \\
 G(x, \eta, t,\mu^2)& =&  \frac{4}{P\cdot n} \int \frac{d \kappa}{2 \pi}\: e^{i x (P\cdot n) \kappa}
 \langle s_2, p_2| n_\alpha G^{\alpha\mu}_{a}(-\kappa n) G_{a \mu}^{\phantom{a \mu}\beta} (\kappa n)n_\beta |p_1,s_1\rangle_{(\mu^2)}
\label{eq:defgluonGPDV}
\end{eqnarray}
\end{subequations}
for  unpolarized ones (parity even operators) and
\begin{subequations}
\label{eq:defGPDA}
\begin{eqnarray}
 \Delta q(x, \eta, t,\mu^2)& =& \int \frac{d \kappa}{2 \pi}\: e^{i x(P\cdot n)\kappa}
 \langle s_2, p_2|\bar{q}(-\kappa n) \, n\cdot\gamma\gamma_5\, q(\kappa n)|p_1,s_1\rangle_{(\mu^2)},
\label{eq:defquarkGPDA} \\
\Delta G(x, \eta, t,\mu^2)& =&  \frac{4}{P\cdot n} \int \frac{d \kappa}{2 \pi}\: e^{i x (P\cdot n) \kappa}
\langle s_2, p_2| n_\alpha G^{\alpha\mu}_{a}(-\kappa n) i \epsilon^{\perp}_{\mu\nu} G_{a}^{\nu\beta} (\kappa n)n_\beta |p_1,s_1\rangle_{(\mu^2)}
\label{eq:defgluonGPDA}
\end{eqnarray}
\end{subequations}
for  polarized ones (parity odd operators).  We note that eight further twist-two quark and gluon transversity GPDs can be defined.
GPDs are quite intricate functions, depending on the momentum fraction
$x$, which is the Fourier conjugate variable of the light-ray distance $\kappa$, the skewness $\eta = -n\cdot\Delta/n\cdot P$, which is the longitudinal momentum fraction in the $t$-channel, and the momentum transfer
squared $t$,  corresponding to transversal degrees of freedom.
Moreover, the GPD evolution w.r.t.~the renormalization scale $\mu$, which is
often equated to the virtuality of the space-like photon, can be perturbatively calculated. The spin content of the in- and out-nucleon state can be parameterized by means of a form factor decomposition,
\begin{subequations}
\label{eq:FtoHE}
\begin{eqnarray}
q &=&
\overline{u}_2\!\left[\frac{\gamma^+}{P^+}  H^q + \frac{  i \sigma^{+\mu}\Delta_\mu}{P^+\,2 M} E^q \right]\! u_1\,,
\quad
\phantom{\Delta}\;\; G =
\overline{u}_2\!\left[\frac{\gamma^+}{P^+}  H^{\rm g} + \frac{  i \sigma^{+\mu}\Delta_\mu}{P^+\,2 M} E^{\rm g} \right]\! u_1\,,
\\
\Delta q &=&
\overline{u}_2\!\left[\frac{\gamma^+ \gamma_5}{P^+} \widetilde{H}^q  + \frac{\Delta^+\, \gamma_5}{P^+\, 2M} \widetilde{E}^q \right]\! u_1\,,
\quad
\Delta G =
\overline{u}_2\!\left[\frac{\gamma^+ \gamma_5}{P^+} \widetilde{H}^{\rm g}  + \frac{\Delta^+\, \gamma_5}{P^+\, 2M} \widetilde{E}^{\rm g} \right]\! u_1\,,
\end{eqnarray}
\end{subequations}
where Dirac spinors $u_i\equiv u(p_i,s_i)$ are normalized as $\overline{u}(p,s)\gamma^\mu u(p,s) = 2 p^\mu$ and $V^+ = n\cdot V$ is the $+$ light-cone
component of a four-vector $V^\mu$.

Basic GPD properties follow from the definitions (\ref{eq:defGPDV}, \ref{eq:defGPDA}) and the light-front wave function (LFWF) overlap representation. \\

\noindent{\textbullet\ \em Reduction to parton distribution functions (PDFs).}\\

\noindent
In the forward limit $p=p_1=p_2$ and $s=s_1=s_2$ the GPDs
(\ref{eq:defGPDV}) and (\ref{eq:defGPDA}) reduce to the common definition of unpolarized quark ($q$), anti-quark ($\overline{q}$) and gluon ($g$)
PDFs as well as the corresponding polarized ($\Delta q$, $\Delta \overline{q}$, $\Delta g$) PDFs \cite{Brock:1994er},
\begin{subequations}
\label{GPD2PDF}
\begin{eqnarray}
q(x,\mu^2) &=& H^q(x\ge 0,\eta=0,t=0,\mu^2)\,,\phantom{--} \qquad  \Delta q(x,\mu^2) = \widetilde{H}^q(x\ge 0,\eta=0,t=0,\mu^2)\,,
\\
\overline{q}(x,\mu^2) &=& -H^q(-x\le 0,\eta=0,t=0,\mu^2)\,, \qquad  \Delta \overline{q}(x,\mu^2) = \widetilde{H}^q(-x\le 0,\eta=0,t=0,\mu^2)\,,
\\
g(x,\mu^2) &=&  \frac{1}{x} H^g(x\ge 0,\eta=0,t=0,\mu^2)\,,\phantom{-} \qquad \Delta g(x,\mu^2) = \frac{1}{x}\widetilde{H}^g(x\ge 0,\eta=0,t=0,\mu^2)\,,
\end{eqnarray}
\end{subequations}
where an additional minus sign appears for unpolarized anti-quarks and an additional $1/x$-factor occurs in the gluonic sector.
\\

\noindent{\textbullet\ \em Symmetry properties.}\\

\noindent
Time reversal invariance together with hermiticity implies that the GPDs (\ref{eq:defGPDV}) and (\ref{eq:defGPDA}) are even functions in the skewness $\eta$.
Furthermore, we can classify  GPDs w.r.t.~symmetry properties under $x\to -x$ reflection. Such (anti-)symmetrized GPDs have definite charge conjugation parity and can be assigned with a signature. Charge-odd (even) quark GPDs $H$ and $E$ as well as charge-even (odd) quark GPDs $\widetilde H$ and $\widetilde E$ are (anti)symmetric functions in $x$ and we assign them with the signature factor $\sigma=-1$ ($\sigma=1$). Gluon GPDs are charge even and they possess signature $\sigma=+1$ ($\sigma =-1$) for GPDs $H$ and $E$ ($\widetilde H$ and $\widetilde E$). Note that in our standard conventions quark and gluon GPDs with the same signature have opposite symmetry properties under $x\to -x$ reflection.
\\

\noindent{\textbullet\ \em Polynomiality of GPD Mellin moments.}\\

\noindent
The $x$-moments of GPDs are given as matrix elements of local twist-two operators, which are contracted with the light-like vector $n^\mu$ and, thus, they are even polynomials in $\eta$ of some certain order. For instance, the lowest moments of quark GPDs give the partonic content of nucleon form factors,
\begin{subequations}
\label{GPDs2FFs}
\begin{eqnarray}
F^N_1(t) &=& \left\{{+2/3 \atop -1/3} \right\} \int_{-1}^1\!dx \, H^{u}(x,\eta,t,\mu^2) +
\left\{{-1/3 \atop 2/3} \right\} \int_{-1}^1\!dx \, H^{d}(x,\eta,t,\mu^2)\,,
\\
F^N_2(t)&=& \left\{{+2/3 \atop -1/3} \right\} \int_{-1}^1\!dx \, E^{u}(x,\eta,t,\mu^2) +
\left\{{-1/3 \atop 2/3} \right\} \int_{-1}^1\!dx \, E^{d}(x,\eta,t,\mu^2)\,,
\end{eqnarray}
\end{subequations}
where the upper (lower) line stays for the proton (neutron) state. Since of the integration, the $\eta$ dependence drops out
and, moreover electromagnetic current conservation ensures the  independence on the factorization scale $\mu$.\\

\noindent
{\textbullet\  \em Support (or spectral) properties.}\\

\noindent
It has been shown by means of the so-called $\alpha$-representation that GPDs possess a spectral representation \cite{Mueller:1998fv}, which is called double distribution representation \cite{Radyushkin:1997ki,Radyushkin:1998es}. Let us emphasize that new insights in this representation arose during time
\cite{Polyakov:1999gs,Belitsky:2000vk,Teryaev:2001qm,Mukherjee:2002gb,Tiburzi:2004mh,Hwang:2007tb,Radyushkin:2013hca}. Nowadays this representation is sometimes quoted for a  quark GPD $F \in \{H,E,\widetilde H, \widetilde E\}$, living in the region $|x|\le 1$, in a most general, however, unspecific form as (here and in the following  the $\mu$-dependence is not indicated anymore)
\begin{eqnarray}
\label{F-DD}
F(x,\eta,t)= \int_{-1}^1\!dy\!\!\int_{-1+|y|}^{1-|y|}\!dz\, \delta(x-y-\eta\, z) \left[f(y,z,t)+ z
\left\{
  \begin{array}{c}
   +\eta\, \delta h(y,z,t) \\
     -\eta\, \delta h(y,z,t) \\
   0 \\
   (1/\eta)\, \delta \widetilde{e}(y,z,t)  \\
  \end{array}
\right\}
\right] \quad \mbox{for}\quad
\begin{cases} H  \\  E\\ H+E, \widetilde H \\ \widetilde{E} \end{cases}
\,,
\end{eqnarray}
with the correspondence $F\in\{H,E,\widetilde H, \widetilde E\} \Leftrightarrow f\in\{h,e,\widetilde h, \widetilde e\}$. As explained below%
\footnote{Since GPD $H+E$ has a standard DD representation, the discussion of GPD $H$ is the same as for GPD  $E$. The GPD $\widetilde E$ might possess some peculiarities, too, which can be treated in an rather analogous manner.}, these representations, which are nothing but a Radon transform, are not {\em unique} and can be {\em reduced} to the `standard' single DD ones with a $D$-term addenda \cite{Polyakov:1999gs},
\begin{eqnarray}
\label{H/E-DD}
F(x,\eta,t)= \int_{-1}^1\!dy\!\!\int_{-1+|y|}^{1-|y|}\!dz\, \delta(x-y-\eta\, z) f(y,z) \pm D(x/|\eta|,t) \quad \mbox{for}\quad
\begin{cases} H  \\  E \end{cases}
\,,
\end{eqnarray}
where the antisymmetric function $D(x/|\eta|)$ is restricted to the central region  $|x| < |\eta|$ and completes polynomiality.
Since GPDs are even functions in $\eta$, the DDs  $(\delta)f(y,z,t)=(\delta)f(y,-z,t)$ are symmetric in $z$.  The DD-representation ensures the polynomiality property of Mellin moments, where the $z \eta$ (or $z/\eta$) proportional term $ \delta h$ is needed if the n$th$  $x$-moment is a polynomial of order $n+1$. Furthermore, a DD with definite signature (or charge parity) has the same parity under $y$-reflection as the corresponding GPD under $x$-reflection.\\

\noindent{\textbullet\ \em Positivity constraints.}\\

\noindent
It has been realized that GPDs  in the outer region, i.e., $|x|\ge |\eta|$,  are constrained by positivity \cite{Martin:1997wy,Pire:1998nw,Radyushkin:1998es,Ji:1998pc,Diehl:2000xz}. Because of the renormalization procedure,
which is rather implicit in the standard minimal subtraction scheme, these constraints might be violated in perturbation theory
at higher order accuracy. Positivity constraints
in a most general form are very intricate \cite{Pobylitsa:2002iu,Pobylitsa:2002vw} and are satisfied in the wave function overlap representation \cite{Pobylitsa:2002vi}. This representation was introduced  in the light-front quantization in \cite{Diehl:2000xz,Brodsky:2000xy}.\\

The representation (\ref{F-DD}) implies that we can define  quark (or gluon) GPDs ($y\ge 0$) and antiquark GPDs ($y\le 0$) with the support $-\eta \le x \le 1$ so that the polynomiality properties hold true, see \cite{Mueller:2005ed}.
The naive partonic interpretation is that in the outer region $\eta \le x \le 1$ a parton is exchanged in the $s$-channel
while in the central region $|x| \le \eta $  a quark-antiquark pair is exchanged in the $t$-channel. Furthermore, one might map the region $-\eta\le x \le 0$ into the region  $0\le x \le \eta$ so that such GPDs have as PDFs the support $0 \le x \le 1$. Such GPDs have definite
charge parity and the polynomiality property might hold only for even (or alternatively odd) Mellin moments.

\subsection{GPD representations and their uses}

In phenomenology various GPD representations are utilized or have been proposed to use. They are based on the DD representation or on the
conformal partial wave expansion (PWE) of GPDs.
The DD-representation (\ref{F-DD}) is together with Radyushkin's DD ansatz (RDDA) rather popular, see below (\ref{RDDA}) in Sec.~\ref{nfPD}. In this representation it is rather obvious (known since one decade and earlier) that GPDs incorporate a {\em duality}, i.e., the central region $|x| \le |\eta|$  and the outer region $|x| \ge |\eta|$  have a cross talk. Note that the term {\em holography} is used by us for GPD models that are so strongly constrained that the  GPD in the ($x,\eta$)-plane can be recovered from its value on the cross-over line (assuming that it is known at $\eta=0$). However, in the DD representation flexible models have not been set up which can be easily done by means of the conformal PWE, which has also some numerical advantages for global GPD fitting. Even if it has {\em not strictly} been  shown in the  mathematical sense for all variations of GPD representations, it is a good starting point to suppose that all these representations are equivalent.

\subsubsection{Double distribution representations}
The DD-representation  is not uniquely defined  and we quoted in (\ref{F-DD}) the most unspecified forms, however, all of them can be reduced to some `standard' one, e.g., as given in (\ref{H/E-DD}).  This statement follows from the considerations in \cite{Belitsky:2000vk} and it can be derived in a more elegant manner  from the view point of a  `gauge' transformation \cite{Teryaev:2001qm} which leaves the GPD invariant,
\begin{eqnarray}
\label{gauge-transform}
f(y,z,t) \to f^\prime(y,z,t)=f(y,z,t) +  \frac{\partial\, z  \chi(y,z,t)}{\partial z},\;\delta f(y,z,t) \to \delta f^\prime(y,z,t) = \delta f(y,z,t) -  \frac{\partial \chi(y,z,t)}{\partial y}.\quad
\end{eqnarray}
This can be easily shown by partial integration for a regular `gauge' transformation, where $\chi(y,z,t)$ is taken to be symmetric in $z$ and vanishes at the support boundary, i.e., $\chi(y,z=\pm (1-y),t)=0$ and $\chi(y=0,z,t)=0$. For a singular `gauge' transformation boundary terms must be taken into account%
\footnote{In this case we can exploit the following equality,
\begin{eqnarray}
\label{gauge-function}
\int_0^1\!dy\!\!\int_{-1+y}^{1-y}\!dz\, \delta(x-y-\eta\, z) \left[\frac{\partial}{\partial z} \,  -  \eta\, \frac{\partial}{\partial y}  -
(1-\eta) \delta(1-y-z) + (1+\eta) \delta(1-y+z) - \eta \delta(y)\right] \tau(y,z)
=0.
\nonumber
\end{eqnarray}
}.

A regular `gauge' transformation can be utilized to verify that the $x$-DD representation,
occurring in a general derivation \cite{Belitsky:2000vk} and utilized for the nucleon case in \cite{Radyushkin:2013hca}, can be transformed to a $(1-x)$-DD representation, arising in a diagrammatical calculation \cite{Hwang:2007tb},
\begin{eqnarray}
\label{DD-E}
E(x,\eta,t) &=& \int_0^1\!dy\!\!\int_{-1+y}^{1-y}\!dz\, \delta(x-y-\eta\, z)\left \{ {x\, b(y,z,t) \atop (1-x) e(y,z,t)}\right\},
\quad e(y,z,t) = -b(y,z,t) + \frac{\partial \chi(y,z,t)}{\partial y},
\nonumber\\
&&\mbox{with}\quad \chi(y,z,t) = \int_0^{y}\! \frac{dw}{1-y}\; b\!\left(\! w,\frac{z (1-w)}{1-y},t\! \right).
\end{eqnarray}
Here, it is assumed that  $b(y,z,t)$ vanishes at $z=\pm (1-y)$ and behaves in the vicinity of $y=0$ as $y^{-\alpha}$ with $\alpha <1$. This
implies that $\chi(y,z,t)$ vanishes at the boundary and can be alternatively calculated from $e(y,z,t)$. However, if one likes to map the $(1-x)$-DD representation to the $x$-DD one, the condition $\chi(y=0,z,t)=0$ could be violated and, thus, one can pick up a boundary term at $y=0$,
\begin{eqnarray}
E(x,\eta,t) &=& x \int_0^1\!dy\!\!\int_{-1+y}^{1-y}\!dz\, \delta(x-y-\eta\, z) \left[ -e(y,z,t) + \frac{\partial \chi(y,z,t)}{\partial y} + \delta(y)\, \chi(0,z,t)\right],
\\
&&\mbox{with}\quad \chi(y,z,t) = \int_{\frac{y}{y+|z|}}^y\! \frac{dw}{y}\; e\!\left(\! w,\frac{z w}{y},t\! \right)
\quad\mbox{and}\quad\chi(0,z,t)  =  \int_{\frac{1}{|z|}}^{1}\! dw\; e(y=0 ,z w,t)\,,
\nonumber
\end{eqnarray}
where it is assumed that $e(y=0 ,z,t)$ exist. Note that the boundary term, proportional to $\delta(y)$, in this new DD contributes only to the highest possible power in $\eta^{n+1}$ for a given odd Mellin moment $x^n$.

Obviously, one might also map the unspecified representation (\ref{F-DD}) for the GPDs $H$ or $E$  in one of the forms (\ref{DD-E}) or in
the more popular form (\ref{H/E-DD}) that is supplemented by the $D$-term \cite{Polyakov:1999gs}. Restricting us to non-negative $y$ and relaxing the condition $\chi(y=0,z,t)=0$ we can, e.g., shift the $\delta h$ addendum for a charge-even (or signature-even) GPD $H$ by utilizing a `singular' gauge
transformation
\begin{eqnarray}
\chi(y,z,t) = -\int_y^{1-|z|}\!\!dw\, \delta h(w,z,t) \quad \mbox{with}\quad  \chi(0,z,t)=  -\int_0^{1-|z|}\!\!dw\, \delta h(w,z,t)
\end{eqnarray}
to a term that is concentrated in $y=0$ or in other words to the $D$-term form (\ref{H/E-DD}) \cite{Belitsky:2000vk,Teryaev:2001qm},
where
\begin{eqnarray}
\label{F-DD-1}
h(y,z,t) \to 
h(y,z,t) - \frac{\partial}{\partial z} z \int_y^{1-|z|}\!\!dw\, \delta h(w,z,t)\,, \;\;
D(x,t) = -\theta(1-|x|)\, x \int_0^{1-|x|}\!\!dw\, \delta h(w,x,t)\, .
\quad
\end{eqnarray}

The DD representation (\ref{F-DD}) also ensures that the GPD can be uniquely extended in the whole $(x,\eta)$-plane
(a consequence of the fact that it is the Fourier transform of an entire analytic function \cite{Mueller:1998fv,Mueller:2005ed}), e.g., restricting $y$ to be non-negative yields a GPD,
\begin{eqnarray}
F(x,\eta,t) = \theta(x+\eta) \omega_F(x,\eta,t) +  \theta(x-\eta) \omega_F(x,-\eta,t)\,,
\label{F(x,eta,t)}
\end{eqnarray}
that lives for $|\eta|\le 1$ in the region $-|\eta| \le x\le 1$, respects polynomiality for even and odd moments, and is
symmetric in $\eta$.
For $|\eta|\ge 1$ we find the representation
\begin{subequations}
\label{F-xspace}
\begin{eqnarray}
\label{F-support}
F(x,\eta,t) = \Theta(x,\eta)\, \omega_F(x,\eta,t) +  \Theta(x,-\eta)\, \omega_F(x,-\eta,t)\,,
\end{eqnarray}
where the support restriction
\begin{eqnarray}
\Theta(x,\eta)\equiv{\rm sign}(1+\eta)\theta\!\left(\!\frac{x+\eta}{1+\eta}\!\right)\theta\!\left(\!\frac{1-x}{1+\eta}\!\right)
\end{eqnarray}
ensures that the polynomiality condition holds true for general $\eta$ values. The
function
\begin{eqnarray}
\omega_F(x,\eta,t) = \frac{1}{\eta}\int_0^{\frac{x+\eta}{1+\eta}}\!dy\, f(y,(x-y)/\eta,t)\,,
\end{eqnarray}
should vanish at $x=-\eta$, however, it is not necessarily analytic at the point $\eta=0$.
\end{subequations}

The representation (\ref{F(x,eta,t)}) implies that the GPD in the central region, given by the function $\omega_F$, can be extended to
the outer region by a symmetrization procedure w.r.t.~$\eta$. One can also extend the GPD from the other region to the central one, e.g.,
by truncation of  the Taylor expansion of `incomplete' GPD  moments \cite{Mueller:2005ed} or as in \cite{Kumericki:2007sa,Kumericki:2008di} by constructing the factorization formulae for double virtual Compton scattering in the Euclidean region by adopting dispersion relation and operator product expansion in the unphysical region (analogous to the classical approach in deep inelastic scattering)  \cite{Chen:1997rc}, or by
utilizing the DD representation \cite{Hwang:2007tb}.

If one likes to find from the GPD the DD-function by an inverse Radon transform,
 \begin{eqnarray}
\label{F_i2f_i}
f(y,z,t)=\frac{-1}{2\pi^2}\!\! \int_{-\infty}^\infty\!\!\frac{dx}{x}\!\! \int_{-\infty}^\infty\!\!\!d\eta\;
\frac{\partial}{\partial x} F(x+y+z\eta,\eta,t)
\;\;\mbox{for}\;\;
F(x,\eta,t) = \int\!\!dy\!\!\! \int\!\! dz\, \delta(x-y-\eta\, z) f(y,z,t)\,,
\nonumber\\
\end{eqnarray}
the GPD  in the entire region is needed. It can be obtained from (\ref{F-xspace}) and can be again expressed by the function $\omega_F$.
For the unspecified form (\ref{F-DD}), it is appropriate to decompose the GPD in a standard piece and a $x$-proportional addenda, which corresponds to a DD-function
$f(y,z,t)+ x \delta f(y,z,t)$ or equivalently  $f(y,z,t) + y \delta f(y,z,t) + \eta z \delta f(y,z,t)$. The inversion formula (\ref{F_i2f_i}) can be applied
for both $f(y,z,t)$ and $\delta f(y,z,t)$ where one can utilize the support properties (\ref{F-support}) and might reshuffle the
$x$-differentiation for numerical evaluations.

\subsubsection{Conformal partial wave expansion}
\label{sec:2-PWE}
GPD representations that are based on the conformal PWE are presented in various forms, too. To derive them, one might start with a  mathematically dual representation in terms of generalized functions (in the mathematical sense), which reads, e.g.,  for quark GPDs,  as
\begin{subequations}
\begin{eqnarray}
F(x,\eta,t,\mu^2) = \sum_{n=0}^\infty (-1)^n p_n(x,\eta) f_n(\eta,t,\mu^2)\,, \quad
f_n(\eta,t,\mu^2) = \int_{-1}^1\!dx\, \eta^n c_n(x/\eta) F(x,\eta,t,\mu^2)\,.
\end{eqnarray}
Here, the notation of \cite{Mueller:2005ed} is used: $c_n(x)= \Gamma(3/2)\Gamma(n+1) C_n^{3/2}(x)/2^n \Gamma(n+3/2)$ are expressed in terms of Gegenbauer polynomials $C_n^\nu(x)$ with
index $\nu=3/2$ of order $n$ and the conformal partial waves
\begin{eqnarray}
p_n(x,\eta) = \frac{\Gamma(n+5/2)}{n! \Gamma(1/2) \Gamma(n+2)}\int_{-1}^1\!du\, (1-u^2)^{n+1}\, \frac{d^n}{dx^n} \delta(x-u\eta).
\end{eqnarray}
\end{subequations}
are up to the normalization nothing but Gegenbauer polynomials that contain the weight $(1-x^2/\eta^2)$, however, they are viewed as generalized functions. The Gegenbauer moments $f_n(\eta)$ of the GPD coincide at $\eta=0$ with the Mellin-moments of the corresponding PDF and they  evolve autonomously in the LO approximation (apart from the quark-gluon mixing in the flavor singlet sector). To employ this representation one has first to map it into the common momentum fraction representation, which can be done by smearing \cite{BelGeyMueSch97}, mapping to forward like PDFs \cite{Shuvaev:1999fm,Nor00}, or by means of the Mellin-Barnes integral \cite{Mueller:2005ed,Kirch:2005tt,ManKirSch05}. Utilizing {\em crossing},  the $t$-channel point of view can be implemented
for GPD moments \cite{Polyakov:1998ze,Ji:2000id} by adopting a SO(3)-PWE \cite{Diehl:2003ny}, where the $t$-channel angular momentum $J$ is the conjugate variable to $1/\eta \sim \cos\theta_{\rm cm}$,
where $\theta_{\rm cm}$ is the scattering angle in the center-of-mass frame of the $t$-channel reaction.  Such a SO(3)-PWE can be implemented in the conformal PWE, yielding the `dual' GPD parametrization in terms of forward-like GPDs \cite{Polyakov:2002wz}. This can be also formulated in terms of a Mellin-Barnes integral representations, where forward-like GPDs and expansion coefficients of conformal moments are related to each other by a Mellin transform.  If the mathematical aspects in such representations are understood \cite{Mueller:2005ed,Kirch:2005tt,Polyakov:2007rw}, it is rather simple to set up flexible GPD models, which are employed in the GPD fitting routine \cite{Kumericki:2011zc}.

\subsubsection{Dissipative framework}
The spectral properties ensure that crossing relations among generalized distribution amplitudes (GDAs) and
GPDs can be easily handled w.r.t.~momentum fraction $x$ and skewness $\eta$. They also ensure that the perturbative QCD results for the deeply virtual Compton scattering and deeply virtual meson production amplitudes can be treated in a dissipative framework. This was realized by O.~Teryaev \cite{Teryaev:2005uj} and was then also explicitly demonstrated in next-to-leading order (NLO) \cite{Diehl:2007jb} and higher twist corrections \cite{Moiseeva:2008qd,Braun:2014sta}.
At the leading twist-two level it is obvious that invariance under boost governs the form of the hard-scattering amplitude $T$ and so the imaginary part of the
amplitude can be written in analogy to convolution formulae in inclusive processes as
\begin{subequations}
\label{dissipative}
\begin{eqnarray}
\Im{\rm m} {\cal F}(\xB,t,\Q^2) = \int_{\xi}^1\! \frac{dx}{x}\,  t(x|\Q^2/\mu^2,\alpha_s(\mu^2)) F(\xi/x,\xi,t,\mu^2)\,,
\end{eqnarray}
where $\xi \simeq \xB/(2-\xB)$ is a Bjorken variable like scaling variable and $t(x)$ is the imaginary part of the hard scattering amplitude $T$. The real part of the amplitude with given signature follows then from the corresponding dispersion relation
\begin{eqnarray}
\Re{\rm e} {\cal F}(\xB,t,\Q^2) = \frac{1}{\pi}\int_0^1\!dx\, \frac{x (1+\sigma)+\xi(1-\sigma)}{\xi^2-x^2 -i\epsilon} \Im{\rm m} {\cal F}\!\left(\!\frac{2x}{1+x},t,\Q^2\!\right) + {\cal C}(t,\Q^2)\,,
\end{eqnarray}
where the subtraction constant ${\cal C}$ can be calculated from the `projection' on the mesonic-like part of the GPD, e.g., for  GPD  $H$ from
the convolution formula
\begin{eqnarray}
{\cal C}(t,\Q^2) = \int_0^1\!\frac{dx}{2}\, T\!\left(\!\frac{1+x}{2}|\Q^2/\mu^2,\alpha_s(\mu^2)\!\right) D(x,t,\mu^2).
\end{eqnarray}
\end{subequations}
In this dissipative framework, apart from the subtraction constant, the central region of the GPD does not enter and so one can ask for information that are
not biased by parton distribution function and form factor modeling. As emphasized in \cite{Kumericki:2008di}, together with {\em GPD-duality} and information from parton distribution functions and
sum rules (form factors and GPD moments from Lattice simulations) such a phenomenological approach offers potentially a much cleaner access to
GPDs.

\subsubsection{Wave function overlap representations}
In the above representations positivity constraints, applying to the outer GPD region,  are not implemented.
Only in the LFWF overlap representation  these conditions can be explicitly implemented. Here, the outer region follows from a diagonal parton LFWF overlap,
while the central region comes from an off-diagonal overlap. If LFWFs are modeled, one can easily break the underlying Lorentz symmetry or one is simply not capable to calculate the off-diagonal overlap. Since parton diagonal overlap representations can be naturally converted in the form of DD representations,  one can utilize {\em GPD-duality} to calculate the full GPD from the  parton diagonal LFWF overlap, however, only for the case that the LFWF respects the underlying Lorentz symmetry \cite{Hwang:2007tb,Mueller:2011bk,Hwang:2012qua}. Note also that `Regge' behavior can be implemented in such simple models from the $s$-channel view \cite{Landshoff:1971xb}, however, the problem to implement a $t$-dependent `Regge'-trajectory in such a manner that positivity holds true by construction is not solved yet.

\subsection{A GPD toy example: photon twist-two GPD}
\label{sec:GPDs-toy}

To illustrate that {\em crossing} and {\em duality} can be employed practically, let us stick to a very simple example, utilized in \cite{Gabdrakhmanov:2012aa},
in which one can explicitly perform all mathematical steps very easily. The  charge-even twist-two photon  GDAs, evaluated in \cite{ElBeiyad:2008ss} to the lowest order accuracy from a box diagram%
\footnote{We neglect here a normalization factor and immediately adopt to the GPD variables $(x,\eta)$, which are related to $(z,\zeta)$, used in \cite{ElBeiyad:2008ss}, by $z=(1-x)/2$ and $\zeta = (1-\xi)/2$.}%
,
\begin{eqnarray}
\Phi_i(x,\eta) = \theta(\eta-x) \varpi_i(x,\eta) + \theta(x-\eta) \varpi_i(-x,-\eta) -\sigma_i \left\{x\to -x \right\}\,,
\end{eqnarray}
live in the region $|x|\le 1$. The unpolarized quark GDA has signature $\sigma_1=+1$ and is entirely determined by the function
\begin{eqnarray}
\varpi_1(x,\eta) = \frac{1+x}{1+\eta}\; \frac{1+\eta -2x}{2}\,,
\end{eqnarray}
while a similar expression holds true for the signature-odd ($\sigma_3=-1$) twist-two  GDA.
{\em Crossing} implies that we can represent the corresponding twist-two photon GPD as
\begin{eqnarray}
\label{GPD-H_i-prel}
H_i(x,\eta) =  \theta(x+\eta)\, \omega_i(x,\eta) + \theta(x-\eta)\, \omega_i(x,-\eta) -\sigma_i  \left\{ x\to -x\right\}\,,
\end{eqnarray}
which also lives in the region $|x|\le 1$, however, is build from the function
\begin{eqnarray}
\label{omega_1}
\omega_i(x,\eta) = \varpi_i(x/\eta,1/\eta)\,, \quad\mbox{e.g.,}\quad  \omega_1(x,\eta) =\frac{1}{2\eta}\frac{x+\eta}{1+\eta}
\left[1+\eta-2x\right]= \frac{x+\eta}{2\eta} - \frac{x}{\eta} \frac{x+\eta}{1+\eta}\,,
\end{eqnarray}
that is obtained from $\varpi_i$ by rescaling $x\to x \eta$ and the inversion  $\eta\to 1/\eta$.
It can be easily established that this representation yields the result of the independent GPD calculation \cite{Friot:2006mm}.

The fact that the GDA/GPD is expressed by only one function $\varpi_i/\omega_i$ implies {\em GPD-duality}, i.e., the cross talk of the central and
outer regions in which the GPD is given by
\begin{eqnarray}
\omega_1(x,\eta) -\omega_1(-x,\eta) =\frac{x(1-\eta )}{\eta (1+\eta )}
\quad\mbox{and}
\quad \omega_1(x,\eta) + \omega_1(x,-\eta) = \frac{1+2x^2-\eta ^2}{1-\eta ^2}-\frac{2x}{1-\eta ^2}\,,
\end{eqnarray}
respectively.  Knowing the function $\omega_1(x,\eta)$, which vanishes at $x=-\eta$, we can immediately construct a GPD that
lives in $-\eta \le x \le 1$. Its even and odd Mellin moments are even polynomials in $\eta$,
\begin{eqnarray}
\label{H1-Mellin}
\int_{-\eta}^1\!dx\, x^n\left[\theta(x+\eta)\, \omega_1(x,\eta) + \theta(x-\eta)\, \omega_1(x,-\eta) \right]=
\frac{2+n-\frac{1}{2}\left[1-(-1)^n\right] \eta ^{1+n}}{(1+n) (2+n)}
-\sum_{k=0}^{n+1}\frac{\left[1+(-1)^k\right] \eta^k}{ (2+n) (3+n)}.
\nonumber\\
\end{eqnarray}
Let us now suppose that we know the GPD only in the outer region and that for these signature-odd GPDs the even moments vanish. One can easily verify that the $x$-symmetric part in the outer region
satisfy polynomiality for odd  moments and includes the highest possible power in $\eta$, i.e., $\eta^{n+1}$. Hence, its extension in the central
region is given by zero. The anti-symmetric part in the outer region does not satisfy polynomiality and it is an algebraic exercise to
find a function in the central region, where an ambiguity w.r.t.~highest possible polynomial in $\eta$ is left. Requiring that the GPD is
continuous at $x=\eta$ provides a unique answer. Such a GPD with definite signature can be defined as for common parton distribution functions
in the region $0\le x \le1 $. For the photon GPD we can quote the odd Mellin moments as
\begin{eqnarray}
\frac{2+n-\eta ^{1+n}}{(1+n) (2+n)}- \sum _{k=0}^{n+1}  \frac{\left[1+(-1)^k\right] \eta ^k}{(2+n) (3+n)} =
\int_{0}^\eta \!dx \frac{x(1-\eta )}{\eta (1+\eta )} x^n +
\int_{\eta}^1 \!dx \left[\frac{1+2x^2-\eta ^2}{1-\eta ^2}-\frac{2x}{1-\eta ^2}\right] x^n,
\end{eqnarray}
which coincide with (\ref{H1-Mellin}).  As we saw the polynomiality condition together with the requirement that the GPD is continues, i.e., no jumps at
the cross-over points $x=\pm \eta$, allows us  to restore  the central GPD region from the outer one. Otherwise if we know the GPD in the central region,
e.g., vanishing at $x=-\eta$ we can construct the outer region. In our example these restorations turns out to be unique for  given signature,
i.e., the completeness of polynomiality, i.e., the so-called `$D$-term' contribution, is associated with both regions.

Let us consider now the DD representation, which we can obtain from an inverse Radon transform (\ref{F_i2f_i}), to illustrate that we can change the GPD  interpretation w.r.t.~$D$-term, however, not the GPD itself. Utilizing the correspondences
\begin{eqnarray}
\left[\frac{(x+\eta)\Theta(x,\eta)}{2\eta} + \eta\to-\eta\right] \Leftrightarrow  \delta(z)+ \eta z \frac{|z|-1}{2|z|} \delta(y)\quad\mbox{and}\quad
\left[\frac{\Theta(x,\eta)}{2\eta}\frac{x+\eta}{1+\eta} + \eta\to -\eta\right] \Leftrightarrow \frac{1}{2} \,,
\nonumber
\end{eqnarray}
where the DD-functions are restricted to the support $0 \le y\le 1-|z|$. The photon GPD, given in terms of the $\omega_1$-function (\ref{omega_1}), has the representation
\begin{eqnarray}
\label{H_1-DD-1}
H_1(x,\eta)=\int_{0}^1\!dy\!\!\int_{-1+y}^{1-y}\!dz\, \delta(x-y-\eta\, z)
\left[\delta(z)+ \eta z\, \frac{|z|-1}{2|z|}\delta(y) -x  \right] - \{x\to -x\}\,.
\end{eqnarray}
Here, both the $x$- and $\delta(y)$ proportional term provide for odd Mellin moments the highest possible power of $\eta$.
After a singular `gauge' transformation (\ref{gauge-transform}), where we set $x=y+ z\eta$  and $\delta h_1 = -1$, the `standard' DD-representation occurs
\begin{eqnarray}
\label{H_1-DD-2}
H_1(x,\eta)&=&  \int_{0}^1\!dy\!\!\int_{-1+y}^{1-y}\!dz\, \delta(x-y-\eta\, z)\, \left[1-2y -2|z|\right] +\theta[x]+ D(x,\eta)- \{x\to -x\},
\nonumber \\
D(x,\eta) &=& -\theta(|\eta|-|x|)\frac{x}{2|\eta|}\left[ 1 + \frac{|\eta|}{|x|} -  2\frac{|x|}{|\eta|} \right]\,,
\end{eqnarray}
which possesses a rather smooth DD in the regular part, a $\eta$ independent $\theta[x]$-term  that results from $\delta(z)$, and the $D$-term
that ensures the completeness of polynomiality.  It can be  verified that both DD-representations (\ref{H_1-DD-1}) and (\ref{H_1-DD-2}) yield
the same GPD that is specified by the $\omega$-function  (\ref{omega_1}). In other words the generalized distributions
(modulus, $\theta$-, $\delta$-, and/or $D$-terms) in the
DD-representation ensure that this function is analytic in $\eta$.

Our photon GPD can be exactly mapped into the space of conformal moments. We calculate the conformal moments  in the outer region,
where {\em GPD-duality} tells us that the non-polynomial part can be thrown away. As explained for Mellin moments, this neglected piece matches for odd moments precisely the contribution coming from the central region. The net result can be quoted in terms of two Gegenbauer polynomials with index $\nu=3/2$,
\begin{eqnarray}
\label{H_1-moments}
H_j &=& \frac{(j+1) (j+2)+2}{(j+1) (j+2)}\, \frac{\Gamma(3/2) \Gamma(j+3)\eta ^{j+1}}{\Gamma(j+5/2)2^{j+1}}\!\left[\!
{_2F_1}\!\left(\!{-j-1,4+j \atop 2}\bigg|\frac{\eta-1}{2 \eta }\!\right)-
{_2F_1}\!\left(\!{-j+1,2+j \atop 2}\bigg|\frac{\eta-1}{2 \eta }\!\right)\!\right],
\nonumber\\
\end{eqnarray}
which we express in terms of hypergeometric functions so that Carlson theorem holds true for positive $\eta$. Hence, the analytical continuation into the complex plane is done and the result together with the diagonal evolution operator (at LO) can be plugged into the Mellin-Barnes integral%
\footnote{Shifting the integration variable $j$ to $j+2$ in the term with the second hypergeometric function improves the convergency of
the integral for $j\to \infty$ and yields an additional subtraction term $-x(\eta^2-x^2)/\eta^3$. Note that this term differs from the $D$-term (\ref{H_1-DD-2}).}. %

We add as a toy example  that if we would take Gegenbauer polynomials with index $3/2$ as SO(3) partial waves we could say
that the $t$-channel angular momentum takes the values $J\in \{j - 1, j + 1\}$ for given conformal spin $j+2$. Furthermore, adopting to
the normalization of parton distribution functions,
{\rm our}  two forward-like GPDs $Q^{\rm our}_0(x)$ and $Q^{\rm our}_2(x)$ in such a `dual' parametrization coincide
with the parton distribution function, simply obtained by an inverse Mellin transform,
$$\frac{2+(j+1) (j+2)}{(j+1) (j+2)(j+3)}\quad \Leftrightarrow \quad Q^{\rm our}_0(x)=-Q^{\rm our}_2(x)= x^2+(1 - x)^2.$$
On the other hand, using the common normalization  of SO(3) partial waves yields
more cumbersome expressions for the forward-like GPDs and can artificially  introduce singularities $Q_2(x)\sim Q_0(x)/x^2$. 
Note, however, that the change of normalization will alter the integral transformation of forward-like GPDs to GPDs or amplitudes.

\section{Status of theory and GPD modeling}
\label{sec:2}

The GPD framework for the description of DVCS and DVMP was set up to  LO accuracy for more than one decade, e.g., in
\cite{Frankfurt:1995jw,Radyushkin:1996ru,Frankfurt:1997fj,Mankiewicz:1997uy,Mankiewicz:1997aa,Mankiewicz:1998kg,Frankfurt:1999xe,Frankfurt:1999fp}, including
the LO evolution equations \cite{Ji:1996nm,Radyushkin:1997ki,Blumlein:1997pi,Balitsky:1997mj,Belitsky:1998gc} that can be also obtained by means of the support extension (rather analogous to {\em GPD duality}) \cite{Dittes:1988xz} from the evolution kernels of meson distribution amplitudes \cite{Lepage:1980fj,EfrRad80,Chase:1980hj,Ohrndorf:1981uz,Baier:1981pm} or from the renormalization
group equation of light-ray operators \cite{Braunschweig:1982mi,Braunschweig:1987dr,Balitsky:1987bk}.
Also GPD models were suggested \cite{Ji:1997gm,Radyushkin:1997ki,Musatov:1999xp} and the first  routines were
written to provide model predictions \cite{Vanderhaeghen:1999xj,Belitsky:2001ns,Freund:2001hd}. This start up worked qualitatively, however, it  was/is not improved as the experimental data base
increased. Hence, this first attempt with relatively rigid GPD models, often to  LO accuracy, was criticized \cite{Kumericki:2008di} and a
GPD fitting framework has been set up \cite{Kumericki:2007sa,Kumericki:2009uq,Lautenschlager:2013uya}. Let us emphasize that GPDs, see (\ref{eq:defGPDV})--(\ref{eq:FtoHE}), are universally defined in the collinear QCD framework, however, they depend on both the light-like vector $n$ and the calculation scheme, i.e., on the considered order in perturbation theory and twist-approximation. Thus,  one should stay inside this framework otherwise the term `universality' loses its meaning. In the following we shortly summarize  what has been calculated in the collinear framework and the models which are utilized.

\subsection{Beyond leading order and leading twist}

The perturbative theory of DVCS and DVMP processes is nowadays available for the leading twist-two approximation
to next-to-leading (NLO) accuracy, which consist of the one-loop corrections to the hard scattering amplitudes for DVCS
\cite{Belitsky:1997rh,Mankiewicz:1997bk,Ji:1997nk,Ji:1998xh,Pire:2011st} and DVMP \cite{Belitsky:2001nq,Ivanov:2004zv}  and the two-loop
corrections to the evolution kernels \cite{Belitsky:1998gc,Belitsky:1999hf}. Apart from the momentum fraction representation, the conformal moments of these quantities are also known as
analytic expressions.  A key ingredient in calculating these corrections was the understanding of conformal symmetry breaking in the perturbative QCD sector \cite{Mueller:1993hg,Mueller:1997ak},
which allows to use for the DVCS hard-scattering amplitudes and evolution kernels known results from deep inelastic scattering.
For a specific factorization scheme in which conformal symmetry is explicitly implemented, the DVCS results in the twist-two sector were given
to NLO and next-to-next-leading order (NNLO) accuracy \cite{Mueller:2005nz,Kumericki:2006xx} and the sizes of QCD radiative corrections were explored in \cite{Kumericki:2007sa}.
While the result for
the hard scattering amplitudes \cite{Belitsky:1997rh} has been confirmed by diagrammatical calculations \cite{Ji:1997nk,Ji:1998xh,Mankiewicz:1997bk,Pire:2011st}, the evolution kernels are so far not diagrammatical evaluated, except
for the  flavor nonsinglet \cite{Dittes:1983dy,Sarmadi:1982yg,Mikhailov:1984ii} and quark transversity \cite{Mikhailov:2008my} sector.
The NLO corrections to the hard-scattering amplitude to DVMP in the non-singlet sector can be simply obtained by the analytic continuation of the pion-form
factor result, given in \cite{Melic:1998qr} and references therein,  where the discontinuity on the real axis is determined by Feynman`s causality prescription. The gluon-quark channel and the pure quark singlet contribution, needed for DVMP of (light) vector mesons, was diagrammatically evaluated
in \cite{Ivanov:2004zv}, where the latter has been recalculated in \cite{Mueller:2013caa}. A consistent presentation of the analytic
results, matching the GPD definitions (\ref{eq:defGPDV})--(\ref{eq:FtoHE}), can be found in \cite{Kumericki:2007sa,Mueller:2013caa}.

The NLO results for time-like DVCS \cite{Berger:2001xd}, diagrammatically calculated in \cite{Pire:2011st}, and DVMP, e.g., exclusive Drell-Yan, processes can be obtained by analytic continuation from the space-like region,
where the factorization scale remains to be positive \cite{Muller:2012yq}. Note that this can be simply seen if one writes the NLO result in terms of convolution integrals and adopts Feynman's causality prescription for the LO result (including a logarithmic modification) \cite{Belitsky:1997rh}. In an analogous manner, one can obtain the NLO coefficients for double DVCS \cite{Guidal:2002kt,Belitsky:2002tf,Belitsky:2003fj}. We add that the NLO result of
gluon transversity for general photon virtualities  is completed, too \cite{Belitsky:2000jk}.

Although no factorization theorems exist at twist-three
level, the explicit LO results \cite{Anikin:2000em,Penttinen:2000dg,Belitsky:2000vx,Kivel:2000cn,Radyushkin:2000ap,Radyushkin:2001fc} and the NLO ones in the flavor non-singlet sector \cite{Kivel:2003jt} do not contradict the hypothesis that factorization holds at this level in DVCS, however, it is expected
that  this will not be true for DVMP amplitudes which might be caused by final state interaction \cite{Collins:1996fb} and shows up at LO as end/cross-over point singularities \cite{Radyushkin:1997ki,Mankiewicz:1999tt}. A closer look to gluon transversity, given in \cite{Kivel:2001qw}, shows that in transverse helicity flip channels factorization is not spoiled at LO accuracy. The twist expansion implies immediately
an ambiguity, namely, the choice of scaling variables depends on the light-cone direction. This ambiguity can be resolved by the evaluation of
twist-corrections beyond the twist-three level. This task turns to be out non-trivial even at LO. The problem in such calculations is the arrangement of
higher twist operators, e.g., quark-gluon-gluon-quark operators, so that a separation of kinematic and dynamic twist-four corrections can be achieved in such a manner that current conservation and Poincar{\'e} co/invariance are restored at the considered twist level.
This has been mastered for the DVCS amplitude to twist-four accuracy for a certain choice of light-like vectors \cite{Braun:2011dg,Braun:2011zr,Braun:2012bg,Braun:2012hq,Braun:2014sta}.

Both the perturbative and twist truncation implies that  the expressions of hadronic observables in terms of  partonic distributions
suffer from some remaining ambiguities.
In particular for the DVCS observables, build from the leptoproduction cross section, these hadron-parton relations are more intricate and depend to some extend on the chosen conventions for the light-like vector. It exist one complete and exact formulae set for
the electroproduction cross section, expressed in so-called helicity dependent Compton form factors (CFFs), that has been given for vanishing electron mass in an analytic form \cite{Belitsky:2012ch}.
This allows easily to express the leptoproduction cross section for any given CFF convention, see examples in \cite{Braun:2014sta}. We add that the inclusion of radiative QED corrections is important for the electroproduction of photons \cite{Vanderhaeghen:2000ws,Bytev:2003qf,Afanasev:2005pb,Akushevich:2012tw}.

\subsection{GPD modeling}
\label{nfPD}

For non-forward parton densities [or skewless GPDs] (nfPDs) modeling is based on phenomenological PDF parametrizations from global fits and sum rules for form factors where it is expected that in future also lattice results for generalized form factors, see \cite{Hagler:2009ni,Bali:2013dpa} and references therein, are sufficiently accurate to be included. Although one may gain some insights in GPDs from such results, it is also clear that one cannot construct GPDs from a (very) limited set of integer moments. The lowest Mellin moments of charge odd quark GPDs yield the nucleon form factors (\ref{GPDs2FFs}).
Furthermore, the $H^{q^{(-)}}$ GPDs at
$\eta=0$ are given by the  valence quark PDFs. Thus, several authors relied on a model ansatz  for the valence quark nfPDs at some typical input scale $\mu\sim 1-2\,\GeV$ and claimed that the momentum fraction dependence for $-t >0 $ can be constrained from form factor data \cite{Guidal:2004nd,Diehl:2004cx,Diehl:2013xca}.

\begin{figure}[t]
\centering
\includegraphics[width=15.5 cm]{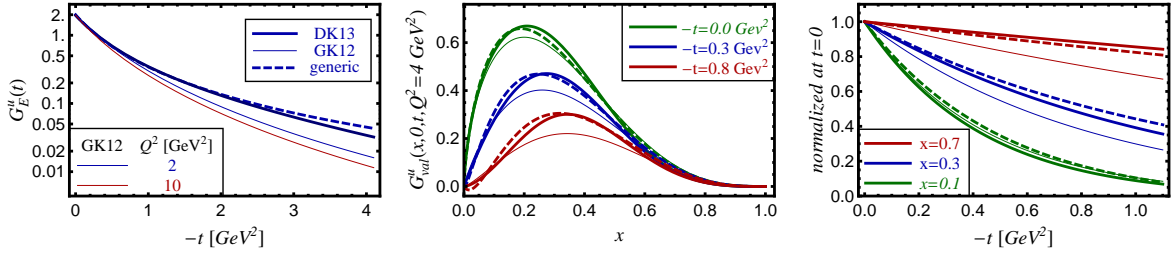}
\caption{Valence $u$-quark form factor (left panel),
valence $u$-quark GPD $G^{u^{(-)}}(x,\eta=0,t,\mu^2=4\, \GeV^2)$ versus $x$ for $t=0$, $t=-0.4\, \GeV^2$, and $t=-0.8\, \GeV^2$ (middle panel) and
the ratio $G^{u^{(-)}}(x,\eta=0,t,\mu^2=4\, \GeV^2)/G^{u^{(-)}}(x,\eta=0,t=0,\mu^2=4\, \GeV^2)$  for $x=0.1$, $x=0.4$, and $x=0.7$ (right panel). The thick solid lines are the default model from \cite{Diehl:2013xca}  with the ansatz (\ref{G-mod2}), the thin solid lines display the results from the {\em GK12} model result, and the dashed lines arise from a generic GPD model that is based on the ansatz (\ref{G-mod1}).
}
\label{fig:G}       
\end{figure}

On the other hand it is clear that such modeling can be trivialized if one works with GPD moments since any phenomenological parametrization of the form factor can be dressed with $j$-dependence. A simple example is the dipole parametrization of the Sachs form factors,
\begin{eqnarray}
G_E^p = \frac{1}{1+\kappa^p} G_M^p = \frac{1}{\kappa^n} G_M^n = \frac{1}{(1-t/m^2)^2}\,,\quad G_E^n =0\,,
\end{eqnarray}
where  $\kappa^p=1.793$ and $\kappa^n = -1.913$ are the anomalous magnetic moments  and  $m_V = 0.84\,\GeV$ is the cut-off mass. The Mellin moments of the `electric' GPD $G= H + (t/4M^2) E$ might be modeled in the simplest version as
$$
\frac{q_j(\mu^2)}{(1-t/m_{j}^2)^2}\quad\mbox{with}\quad m^2_{0} =  0.84^2\,\GeV^2 \quad \mbox{and}\quad\lim_{j\to \infty} m^2_{j} =\infty,
$$
where the latter condition ensures that the $t$-dependence flatter out at large $j$ and, consequently, also at large $x$. It might be considered as appealing
to model the $t$-dependence by means of Regge-trajectory. We might set $m_j^2= (j+1-\alpha)/\alpha^\prime$ and find with $\alpha\approx 1/2$ and
$\alpha^\prime \approx 4/5$ for $j=0$ approximately the quoted value for $m_V$. The Mellin transform provides then the corresponding GPD
\begin{eqnarray}
\label{G-mod1}
G(x,\eta=0,t,\mu^2) &=& \frac{ \Gamma(2-\alpha +\beta) \Gamma(1+\alpha^\prime t) }{\Gamma(1-\alpha) \Gamma(1+\beta +\alpha^\prime t)}x^{-\alpha }
\Bigg[
\left(1-\left\{S_1(\beta +\alpha^\prime t)-S_1(\alpha^\prime t)+\ln x\right\}\alpha^\prime t \right) x^{\alpha^\prime t}
\\
&&\qquad\qquad\qquad\qquad\qquad - \frac{\Gamma(1+\beta +\alpha^\prime t)\, x}{\Gamma(\beta) \Gamma(2+\alpha^\prime t) (1+\alpha^\prime t)}\,
{_4F_3}\!\left(\!{2,1-\beta ,1+\alpha^\prime t,1+\alpha^\prime t\atop 1,2+\alpha^\prime t,2+\alpha^\prime t}\Big|x\!\right)
\Bigg]
\nonumber
\end{eqnarray}
The functional form of this result differ from a factorized exponential ansatz,
\begin{eqnarray}
\label{G-mod2}
G(x,\eta=0,t,\mu^2) = q(x,\mu^2) \exp\left\{f(x) t\right\} + e(x,\mu^2) \exp\left\{g(x) t\right\},
\end{eqnarray}
which are utilized, e.g., in Refs.~\cite{Guidal:2004nd,Diehl:2004cx,Diehl:2013xca}.  We can clearly state that form factor sum rules do not allow to pin down the $x$-shape of GPDs. Utilizing PDF information in addition and assumption about the $t$-behavior at the $x$-space boundaries, sum rules might be able to constrain the interplay of $x$- and $t$-dependence to some certain extend. In Fig.~\ref{fig:G} it is illustrated that under such circumstances similar PDF shapes yield similar results for the nfPD. Indeed the predictions of the model (\ref{G-mod1}) [thick lines] and the default fit (\ref{G-mod2}) [thin lines] from \cite{Diehl:2013xca}
for the $x$-dependence (left panel)  as well as for the $t$-dependence (right panel) for the valence $u$-quark GPD $G^{u^{(-)}}$ are very similar. To mimic the $x$-shape of the NLO PDF parametrization
\cite{Alekhin:2012ig}, which is used in \cite{Diehl:2013xca}, we set  $\alpha=0.3$ and $\beta=3$ in (\ref{G-mod1}).

Skewing GPDs is often done in the DD-representation with RDDA, improved by the $D$-term \cite{Polyakov:1999gs} and pion-pole  contribution \cite{Mankiewicz:1998kg,Frankfurt:1999fp} and also a change of the DD representation (not of the DD) is utilized \cite{Radyushkin:2013hca}. RDDA is inspired from tree-level diagrams (or one may say from diquark models) and reads for $t=0$ at some
given input scale as
\begin{eqnarray}
F(x,\eta,t=0) =  \int_0^1\!\!dy \int_{-1+y}^{1+y}\!dz\,\delta(x-y-z\eta)\, \frac{F(y,\eta=0,t=0)}{1-y}\, \frac{\Gamma(3/2 + b)}{\Gamma(1/2) \Gamma(1 + b)}\left[1-\frac{z^2}{(1-y)^2}\right]^b.\quad
\label{RDDA}
\end{eqnarray}
Adopting the PDF pararametrization $F(y,\eta=0,t=0)= n\, y^\alpha (1-y)^\beta$, one can show that this model is {\em holographic} in the sense that one can restore the GPD from the PDF and its values on the cross-over line and  so this ansatz incorporates strong constrains \cite{Kumericki:2010fr}. Unpleasant for the phenomenology of DVCS/DVMP is the fact that the small-$x$ and large-$x$ behavior of the GPD on the cross-over line,
\begin{eqnarray}
F(x,x,t=0) = \frac{n\,\Gamma(3/2+b) \Gamma(1+b-\alpha)}{2^{-2 b-1}\Gamma(1/2) \Gamma(2+2 b-\alpha)}
\left(\!\frac{2x }{1+x }\!\right)^{-\alpha }\! \left(\!\frac{1-x }{1+x }\!\right)^{\beta}\!
  {_2F_1}\!\left(\!{1+b,1-\alpha +\beta\atop 2+2 b-\alpha}\bigg|\frac{2 x }{1+x }\!\right)\!,\quad
\end{eqnarray}
are controlled by the profile parameter $b$. In the small $x$-region one finds the following skewness ratio
\begin{eqnarray}
r(x,t=0) \stackrel{x \sim 0}{\approx} \frac{2^{2 b+1}\Gamma(3/2+b) \Gamma(1+b-\alpha)}{2^{\alpha}\Gamma(1/2) \Gamma(2+2 b-\alpha)}\quad\mbox{for}\quad
r(x,t=0)= \frac{F(x,x,t=0)}{F(x,0,t=0)}\,,
\label{r(x,t=0)}
\end{eqnarray}
which is for typical Regge behavior $\alpha>0$ and positive $b \gtrsim \alpha$  larger than one and takes in the limit $b\to \infty$, in which the skewness dependence
is removed, the value one.
It is sometimes believed, coming from a wrong mathematical understanding of the conformal PWE \cite{Kumericki:2009ji},
that this ratio is fixed by setting $b=\alpha$.  In the large-$x$ region the skewness ratio is divergent for $b<\beta$ and
finite for $b\ge\beta$,
\begin{eqnarray}
r(x,t=0)\stackrel{x \sim 1}{\approx}
\frac{2^{b+1}\Gamma(\frac{3}{2}+b)}{ \Gamma(1/2) \Gamma(1+b)}
 \left[
\frac{\Gamma(1-\alpha+b) \Gamma(\beta-b)}{\Gamma(1-\alpha +\beta)} (1-x)^{b-\beta}
+
\frac{\Gamma(b-\beta) \Gamma(1+b)}{2^{\beta-b} \Gamma(1+2 b-\beta)}\right].
\end{eqnarray}

Neglecting evolution, in a LO description of DVCS/DVMP data  one needs only to model the GPD on the cross-over line.
A very simple example that is inspired from a diquark GPD model reads as
\begin{equation}
F(x,x,t) =  \frac{n\, r}{1+x} \left(\!\frac{2x }{1+x }\!\right)^{-\alpha(t) }\! \left(\!\frac{1-x }{1+x }\!\right)^{b}
\left(1-\frac{1-x}{1+x} \frac{t}{M^2}\right)^{-p}\,.
\label{F(x,x,t)}
\end{equation}
Here, the normalization factor $n$ is fixed from a reference PDF parametrization while the independent skewness ratio parameter $r$, which controls the normalization at small $x$, overcomes the rigidity of the RDDA.
The $t$-dependence is contained in the linear Regge trajectory $\alpha(t)=\alpha(0)+ \alpha^\prime t$ and the residual $t$-dependence is a $p$-pole with a $x$-dependent cut-off mass. As expected from theoretical arguments and in agreement with phenomenological findings, the $t$-dependence dies out for $x\to 1$.

If one includes evolution, higher order corrections, and/or higher twist contributions, one has to model the whole GPD.
If it is ensured that the full GPD can be consistently reconstructed, one can  employ the dissipative framework and so it is sufficient to model the GPD
in the outer region. As stated in Sec.~\ref{sec:2-PWE}, the parametrization of GPDs in terms of (conformal) moments that are expanded in terms of SO(3)-partial waves (PWs) has several advantages.  The ansatz, which is used for the description of
small-$\xB$ data, is given in terms of three SO(3)-PWs $\hat{d}_J(\eta)$ for $J\in \{j-3,j-1,j+1\}$, normalized to one for $\eta=0$,
\begin{eqnarray}
F_j(t) = f^{j+1}_j(t)\, \hat{d}_{j+1}(\eta)+ \eta^2 f^{j-1}_j(t)\, \hat{d}_{j-1}(\eta) + \eta^4 f^{j-3}_j(t)\, \hat{d}_{j-3}(\eta)\,,
\label{F_j(t)}
\end{eqnarray}
see photon GPD example (\ref{H_1-moments}). This flexible ansatz with two free skewness parameters $s_2$ and $s_4$, introduced as $f^{j+1-k}_j \propto s_k f^{j+1}_j$, allows to control
both the normalization of the  GPD on the cross-over line and
their evolution (crucial for data description). The leading SO(3)-PW amplitude
$$ f^{j+1}_j(t=0) \sim  \frac{\Gamma(1-\alpha+j)\Gamma(1+\beta)}{\Gamma(2+j-\alpha+\beta)} +\cdots $$
at $t=0$ might be fixed by a PDF parametrization or alternatively by fits to small-$\xB$ deep inelastic scattering (DIS) data.
The $t$-dependence is  introduced via a `Regge' trajectory and a residual $t$-dependence. With the truncated PWE (\ref{F_j(t)})
the GPD behavior in the large-$x$ region
will be determined by the adopted SO(3)-PWs. To overcome this rigidity in a simple manner,
one may employ  `effective' PWs.

Finally, we mention that although the theory is worked out at NLO its implementation in computer codes at this level, e.g., for a fitting routines, has so far not be satisfactorily mastered in the momentum fraction representation. The main difficulty in this representation is to have at hand flexible GPD models. Note that it has been shown to LO accuracy that for fixed GPD models fast and robust numerical routines can be set up by means of two dimensional $x$-$\Q^2$ grids ($\xi$ and $t$ are fixed) \cite{Vinnikov:2006xw} and that an older NLO code \cite{Freund:2001hd}, based on the RDDA, is compatible with  DVCS data from
the DESY collaborations. It is part of the Monte Carlo routine MILOUS, which is tuned to H1/ZEUS data. In a dissipative framework one can potentially adopt the grid technology which was worked out for inclusive processes. Alternatively, a fitting framework was set up that is based on the Mellin-Barnes integral technique \cite{Kumericki:2007sa,Kumericki:2009uq} and is presently extended to global DVCS/DVMP fitting at NLO \cite{Mueller:2013caa}.

\section{Uses of GPDs in phenomenology}
\label{sec:4}

The key application of GPDs is the description of DVCS and DVMP processes. In LO approximation the amplitude of these processes is proportional to
a convolution integral which we generically write for amplitudes with definite charge parity or signature $\sigma \in \{+1,-1\}$  as
\begin{eqnarray}
{\cal F}_M(\xB,t,\Q^2) \propto\!\!\! \sum_{p=u,d,\cdots}\!\!\!\!\! C^p_M\int_{-1}^1 dx \left[\frac{1}{\xi-x-i\epsilon} - \frac{\sigma}{\xi+x-i\epsilon}\right]F^p(x,x,t,\Q^2)\,,
\label{conv-LO}
\end{eqnarray}
where $C^p_M$ are build from quark charges and Clebsch-Gordon coefficients. In the dissipative framework, see discussion around (\ref{dissipative}),  one can name the degrees of freedom that can be accessed to LO accuracy. Namely, in this approximation one  accesses the GPD on the cross-over line, i.e.,
\begin{eqnarray}
{\cal F}_M(\xB,t,\Q^2) \propto\!\!\!  \sum_{p=u,d,\cdots}\!\!\!\!\! C^p_M \left[ i \pi F^p(\xi,\xi,t,\Q^2) + {\cal P}\!\! \int_0^1 dx \frac{x(1+\sigma)+(1-\sigma)\xi}{\xi^2-x^2} F^p(x,x,t,\Q^2) + {\cal C}^p(t,\Q^2)\right]\!\!.
\nonumber\\
\label{conv-DR}
\end{eqnarray}
where $\cal P$ denotes the principal value description and ${\cal C}^p$ is a subtraction constant. The subtraction constant appears for the GPDs $H$ and $E$ with different signs and as explained above it can be expressed by the $D$-term. The dispersion relation (\ref{conv-DR}) can be utilized as a sum rule to access the GPD on the cross-over line (and the subtraction constant) also outside of the experimentally accessible phase space. The reader can find a more detailed discussion in \cite{Kumericki:2008di}.

So far in phenomenology most authors concentrated on the DVCS process, which is theoretically considered  as clean,
however, has a rather intricate azimuthal angular dependence. This lepton charge odd process appears besides the lepton charge even Bethe-Heitler process in the leptoproduction of a photon and is described by twelve complex valued helicity amplitudes or Compton form factors. It has been realized in \cite{Belitsky:2001ns} that a complete measurement is in principle possible, which allows to access all twenty-four sub-CFFs (real and imaginary parts of twelve CFFs). Strategies to analyze and interpret DVCS observables have been utilized and they are based on
\begin{itemize}
\item[$\bullet$] GPD model predictions, essentially based on RDDA, versus measurements \cite{Polyakov:2008xm,Kroll:2012sm}.
\item[$\bullet$] Local sub-CFF estimates (least square fits \cite{Guidal:2008ie,Guidal:2009aa,Guidal:2010de,Guidal:2010ig}, neural networks \cite{Kumericki:2011rz}), regression and random variable maps \cite{Kumericki:2013br}.
\item[$\bullet$] Dispersion relation fits, where the spectral function and subtraction constants are modeled \cite{Kumericki:2009uq}.
\item[$\bullet$] Flexible GPD model fits, essentially by utilizing models for GPD moments \cite{Kumericki:2009uq,Kumericki:2011zc,Aschenauer:2013qpa,Kumericki:2013br,Lautenschlager:2013uya}.
\end{itemize}

GPD predictions are often based on RDDA, see (\ref{RDDA}) and discussion around.  The Vanderhaeghen, Guichon, and Guidal (VGG) code \cite{Vanderhaeghen:1999xj} provides LO predictions where the  GPD models are adopted from \cite{Goeke:2001tz} and \cite{Guidal:2004nd} in dependence on a chosen set of phenomenological PDFs, obtained from global fitting. The Goloskokov and Kroll model \cite{Goloskokov:2005sd,Goloskokov:2006hr,Goloskokov:2007nt,Goloskokov:2008ib,Goloskokov:2009ia} uses fixed values for the profile parameters and the $t$-dependence is implemented as described in Sec.~\ref{nfPD}. However, these two RDDA based models implement the partonic angular
momenta $J_u/J_d$ differently. In VGG they are implemented via sum rules, using a invisible $\delta(x)$ term, \cite{Goeke:2001tz} while in GK models
the valence angular momenta are fixed by the anomalous magnetic nucleon momenta and the momentum fraction shape of the valence GPDs $E$ and the
sea quark and gluon angular momenta are considered as free parameters. Consequently, the usage of the free  $J_u/J_d$ parameters in VGG requires some
caution. It became clear for one decade that such models capture qualitative features
of DVCS data, however, they are not able to accurately describe them at LO, in particular, they overshoot the beam spin asymmetry
measurements of HERMES and CLAS collaborations and the DVCS cross section measurements of the H1 and ZEUS collaborations. Note that the beam spin asymmetry
measurements from HERMES with a recoil detector are slightly higher in their absolute values than the previous results, obtained with the missing mass technique and so the new beam spin asymmetry data are getting compatible with RDDA based model predictions. Nevertheless, one can state that in a global LO analysis such models cannot describe the data on a quantitative level%
\footnote{This conclusion was already given for some time by means of another RDDA based model \cite{Belitsky:2001ns}, however, it has not necessarily to
be true at NLO. Note also that inconsistent GPD models (breaking of polynomiality \cite{Freund:2003qs} and neglecting the `pomeron' components) are used to describe DVCS measurements. }.

\begin{figure}[t]
\centering
\includegraphics[width=10cm]{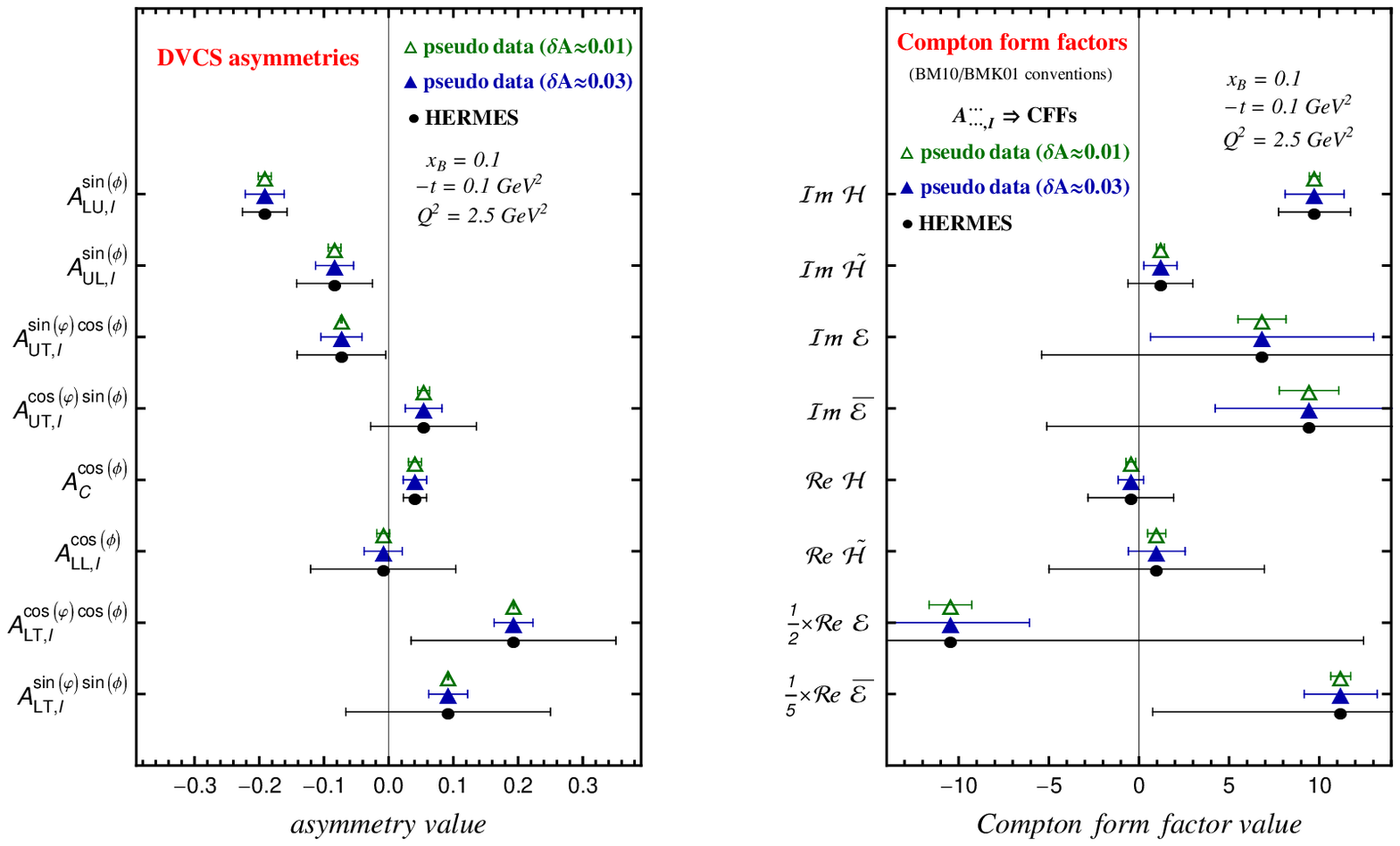}\hspace{4mm}
\includegraphics[width=5cm]{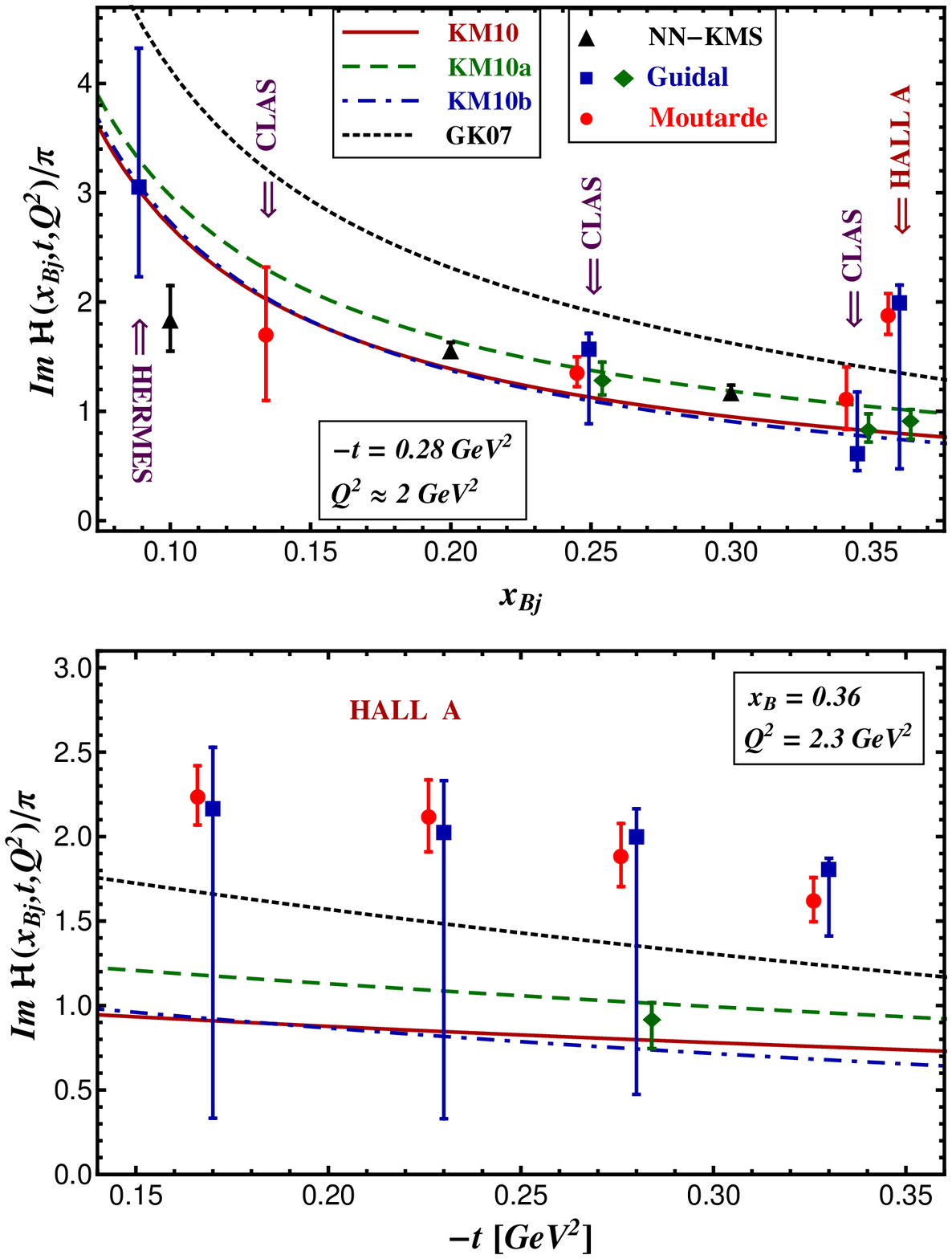}
\caption{
Left: two projections with total errors $0.01 \lesssim \delta A \lesssim 0.03$ (filled triangles) and $\delta A \approx 0.01$
(empty triangles) of charge odd asymmetries (left) for $\xB = 0.1$, $t = −0.1\, \GeV^2$, and $\Q^2 = 2.5\, \GeV^2$
are compared with HERMES measurements (solid circles). Middle: the resulting CFFs from a
one-to-one map, based on the twist-two dominance hypothesis.
Right: $\Im {\rm m} H/\pi$ obtained from different strategies:  DVCS fits dashed (dashdotted, solid) curve excludes
(includes) Hall A data from {\em KM10a} ({\em KM10b}, {\em KM10}) \cite{Kumericki:2011zc},
{\em GK07} model prediction (dotted), seven-fold sub-CFF fit \cite{Guidal:2008ie,Guidal:2009aa}  with boundary
conditions (squares), sub-CFF ${\cal H}$, $\widetilde{\cal H}$ fit \cite{Guidal:2010ig} (diamonds),
smeared conformal partial wave model fit \cite{Moutarde:2009fg} within  GPD $H$ (circles). The triangles result from a
neural network fit \cite{Kumericki:2011rz}.
Figures are taken from \cite{Kumericki:2013br} and  \cite{Kumericki:2011zc}.
}
\label{fig:HERMES}       
\end{figure}
At HERA both kinds of lepton beams were available and so the HERMES collaboration provided an almost (over)complete measurement of {\em thirty-four} DVCS asymmetry harmonics in twelve $\{\xB,t,\Q^2\}$-bins. This would allow in principle to map the asymmetry measurements  to the  {\em twenty-four} dimensional space of sub-CFFs. So far this mapping of random numbers has been only performed under the assumption that only eight twist-two related sub-CFFs contribute. The result is shown in Fig.~\ref{fig:HERMES} for one bin.  The HERMES data were also analyzed by means of regression \cite{Kumericki:2013br} and by truncating the space of sub-CFFs within neural networks \cite{Kumericki:2011rz} and least square fits \cite{Guidal:2009aa,Guidal:2010de}. The results from all these methods
are very similar: the nonzero sub-CFF $\Im{\rm m}{\cal H}$ and the small sub-CFFs  $\Re{\rm m}{\cal H}$ and $\Im{\rm m}\widetilde{\cal H}$ are
well constrained, while all other ones are noisy and compatible with zero. Decreasing the errors of all asymmetries on the level of present beam spin asymmetry measurements would allow to access the CFF ${\cal E}$.

In JLAB experiments the number of measured observables
is presently restricted to unpolarized  and longitudinal polarized proton observables with a polarized electron beam.   Hence, the local extraction of CFFs can be only done by utilizing assumptions or model constraints. Taking only beam spin asymmetry measurements from the CLAS collaboration allows to
provide constraints on sub-CFF  $\Im{\rm m}{\cal H}$, which depend on assumptions. A rather similar  situation occurs for the HALL A cross section measurements at an unpolarized proton with polarized electron beam. The dependence on the constraints is clearly visible in the lower right panel in Fig.~\ref{fig:HERMES}, where the twist-two dominance hypothesis (squares) \cite{Guidal:2010ig}, combined with model constraints, is compared with the ${\cal H}$-dominance hypothesis (circles) \cite{Moutarde:2009fg}. Obviously, the size of propagated errors depends on the assumptions. Only the measurement of new observables can yield an
improvement, which is illustrated by the diamond that takes also into account longitudinal proton spin asymmetry measurements from CLAS \cite{Guidal:2010ig}.

H1 and ZEUS collaborations measured the unpolarized DVCS cross section in the small-$\xB$ region, where the interference term could be safely neglected and the Bethe-Heitler cross section was subtracted, \cite{Adloff:2001cn,Aktas:2005ty,Aaron:2007ab,Aaron:2009ac,Chekanov:2003ya, Chekanov:2008vy} as well as the electron charge asymmetry \cite{Aaron:2009ac}. In this small-$x$ region one expects that and that the `pomeron' behavior of
GPDs $H$ and $E$ imply the leading contributions and that   apart from the
effective `pomeron' trajectory no other $\{x,t\}$ correlations appear.  The collider data can be described with flexible GPD models at LO, NLO, and NNLO, which are modeled in terms of conformal moments (\ref{F_j(t)}). While in a LO description the $r$-ratio, defined in (\ref{r(x,t=0)}), for sea quarks is approximately one while for gluons it is considerable smaller. Beyond the LO the sea quark and gluon $r$-ratio is considerable larger  and compatible with the claim \cite{Shuvaev:1999ce,Martin2009zzb}, i.e., $r^{\rm sea} \sim 1.6$ and $r^{\rm G} \sim 1$. Present data do not allow to access the GPD $E$, which is in the unpolarized cross section   suppressed by $t/4 M_p^2$. Hence, such GPD fits provide the $t$-dependence of $H(x,x,t,\Q^2)$ and allow for a model dependent access of the GPD
$H(x,0,t,\Q^2)$ and, consequently, after Fourier transform for a probabilistic interpretation in impact parameter space.

\begin{figure}[t]
\centering
\includegraphics[width=15.2cm]{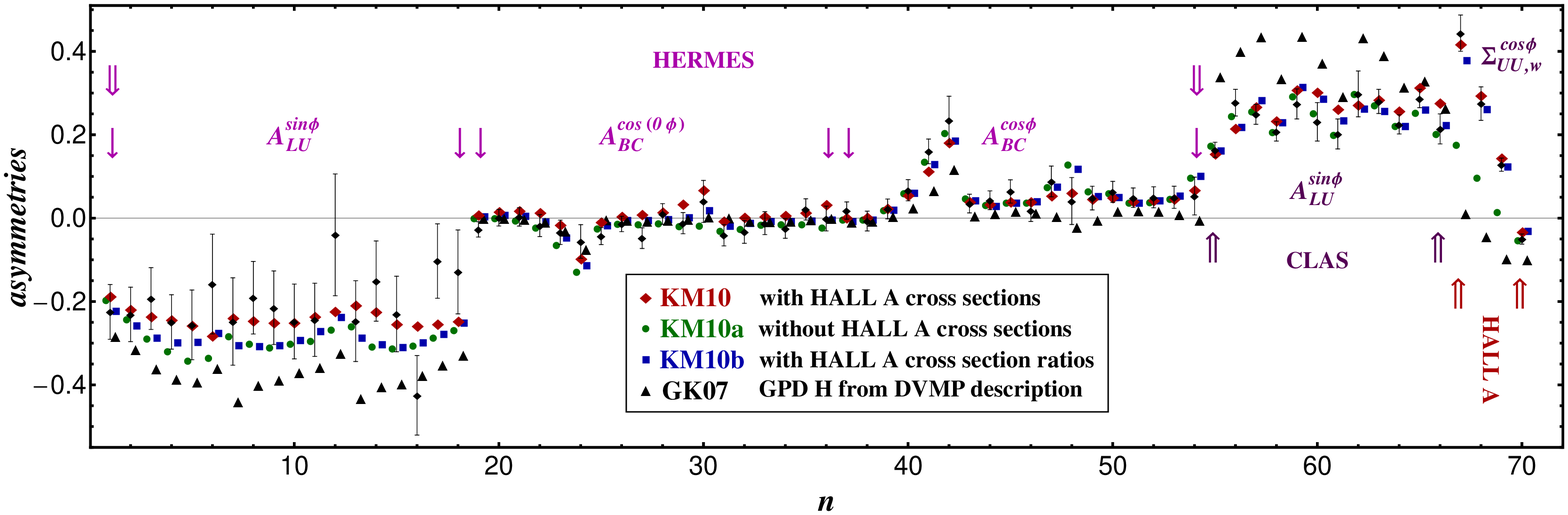}
\caption{Experimental  measurements of DVCS harmonics for fixed target kinematics (error bars) labeled by data point number $n$: beam spin asymmetry ($A^{\sin\phi}_{\rm LU}$, 1-18),  beam charge asymmetry ($A^{\cos(0\phi)}_{\rm BC}$, 19-36 and
$A^{\cos\phi}_{\rm BC}$, 37-54)
from \cite{Airapetian:2009aa}; beam spin asymmetry ($A^{\sin\phi}_{\rm LU}$, 55-66)
and unpolarized cross section ($\Sigma_{\rm UU,w}^{(\cos \phi}$, 67-70) are derived from \cite{Girod:2007aa}
and \cite{Munoz_Camacho:2006hx}. Model results  are pinned down by fits {\em KM10a} without (circles, slightly shifted to the l.h.s.) and
 {\em KM10b} (squares, slightly shifted to the r.h.s.) and  {\em KM10} (diamonds) with Hall A data as well as given by a RDDA based GPD $H$ model prediction {\em GK07}  (triangles-down, slightly shifted to the r.h.s.)~\cite{Goloskokov:2007nt}.
}
\label{fig:FixTar}
\end{figure}
Global GPD fits, restricted to LO  and twist-two accuracy, of DVCS data have been given so far in the dissipative framework \cite{Kumericki:2009uq} or with hybrid models \cite{Kumericki:2011zc} in which the valence content is described by a flexible GPD on the cross-over line (\ref{F(x,x,t)}) and sea quark as well as gluon GPDs have been implemented via conformal moments (\ref{F_j(t)}). The GPD models are designed to described unpolarized proton data and depend on {\em fifteen} parameters which allow to control the skewness- and $t$-dependence, the $\Q^2$-dependence in the small-$x$ region, as well as subtraction constants for the real part of CFFs. In such fits the cross section measurements from HALL A are neglect ({\em KM09a},{\em KM10a}), taken into account as $\phi$-harmonics  ({\em KM09},{\em KM10}, {\em KMM13}) or as their ratios ({\em KM09b},{\em KM10b}). The quality of the fits for unpolarized proton data is $\chi^2/{\rm d.o.f.}\approx 1$ while for polarized proton target the {\em KMM13} fit has $\chi^2/{\rm d.o.f.}\approx 1.6$. This tension might be implied by the description of HALL A cross section measurements with simple GPD models in the twist-two sector. The results of the fits
are shown in Fig.~\ref{fig:FixTar} and compared with a model prediction (triangles-down) \cite{Kumericki:2011zc,Kroll:2012sm}%
\footnote{The predictions are shown by taking into account the dominant GPD $H$, where the remaining three GPDs induce only a slight change.
It is also not entirely clarified why this GPD model also describes H1 and ZEUS data \cite{Meskauskas:2011aa,Kroll:2012sm} while all other consistent GPD models that are based on RDDA and are compatible with DIS data predict too large DVCS cross sections  at  LO accuracy.},
which illustrates the typical problems with beam spin asymmetry and HALL A cross section measurements, discussed above.

\begin{figure}[t]
\centering
\includegraphics[width=15.2cm]{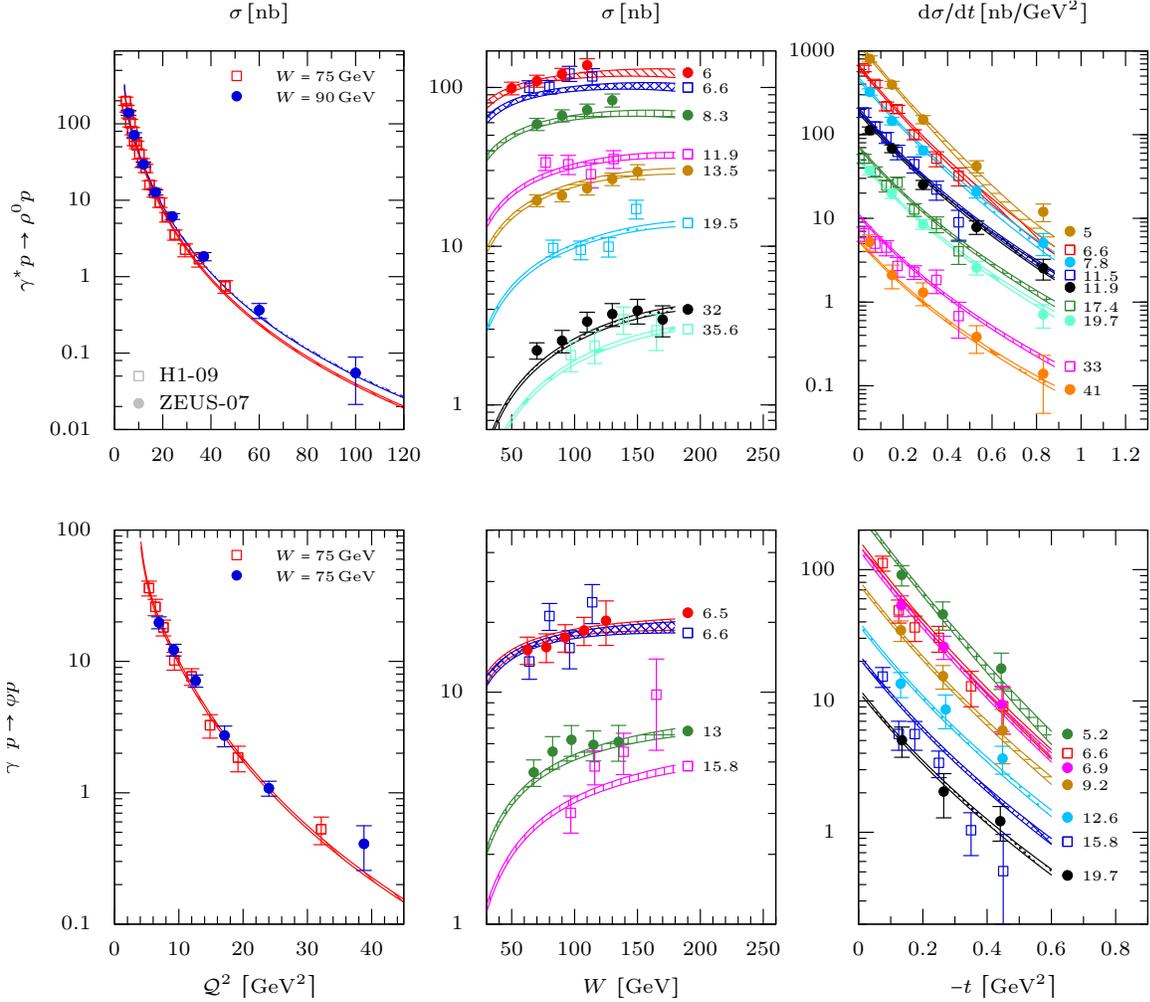}
\caption{
Total cross section vs. $\Q^2$ (left) and $W$ (middle) as well as differential one vs. $−t$ (right) of
DV$\!\rho^0$P (top) and DV$\!\phi$P (bottom) measurements from H1 \cite{Aaron:2009xp} and ZEUS \cite{Chekanov:2005cqa,Chekanov:2007zr}. Curves arise from a
simultaneous GPD estimate at NLO, where numbers next to them are the $\Q^2$ values in units of $\GeV^2$. Figure taken from \cite{Lautenschlager:2013uya}.
}
\label{fig:ColKin}
\end{figure}

Various model predictions have been given for DVMP processes to LO accuracy where transverse degrees of freedom were taken into account.
Most of these predictions use GPDs that are based on  RDDA and include transverse degrees of freedom in the hand-bag approach (corresponding to the LO diagrams that appear in the collinear approach). In this approach DVMP measurements of (light) vector mesons \cite{Vanderhaeghen:1999xj,Goloskokov:2005sd,Goloskokov:2006hr,Goloskokov:2007nt,Goloskokov:2008ib} can be described apart from the
very large $\xB$-region. Here, it was suggested to add a large real part to the amplitudes \cite{Guidal:2007cw}, see also \cite{Hyde:2011ke}, where another possibility to add in first place a large imaginary part has not been explored \cite{Hwang:2007tb}. DV$\pi^+$P data can be described
in terms of the hand-bag approach \cite{Goloskokov:2009ia} or the collinear factorization approach \cite{Bechler:2009me}.
A systematic analysis of these processes in the collinear factorization framework has been started recently, where one should go beyond
the LO level \cite{Bechler:2009me,Meskauskas:2011aa}. In fact one can very well describe for $\Q^2 > 4 \GeV^2$ the H1 and ZEUS measurements of DVCS and DVMP, including the DIS data, at NLO \cite{Lautenschlager:2013uya}. This is shown for the D$\rho^0$MP and  D$\phi$MP data in Fig.~\ref{fig:ColKin}, where the description of DVCS and older DIS data is unproblematic and has the same good quality as in \cite{Kumericki:2009uq,Kumericki:2011zc}.

\section{Conclusions}
\label{sec:LFWFs}

The richness of information, contained in GPDs, make them most valuable sources to reveal the longitudinal and transversal
distributions of partons.  On the other hand their intricate variable dependence complicates their access from experimental measurements
and one should clearly stay that addressing the main goals in GPD phenomenology, i.e., nucleon spin decomposition and imaging,
will be based on GPD modeling. Obviously, to get a partonic interpretation one must first describe the experimental data, which is only possible with flexible GPD models and a global fitting procedure while so far model predictions work only to some certain extend.
Based on the present theory, it is a straightforward procedure to improve the GPD framework of global fitting, developed in \cite{Kumericki:2007sa,Kumericki:2009uq,Mueller:2013caa},
where in future also information on elastic and gravitational (or generalized) form factors  will be employed. Technically, the Mellin-Barnes integral representation allows for a simple implementation of such constraints and, moreover, numerically efficient and robust routines can be written.

\begin{figure}[t]
\centering
\includegraphics[width=13cm]{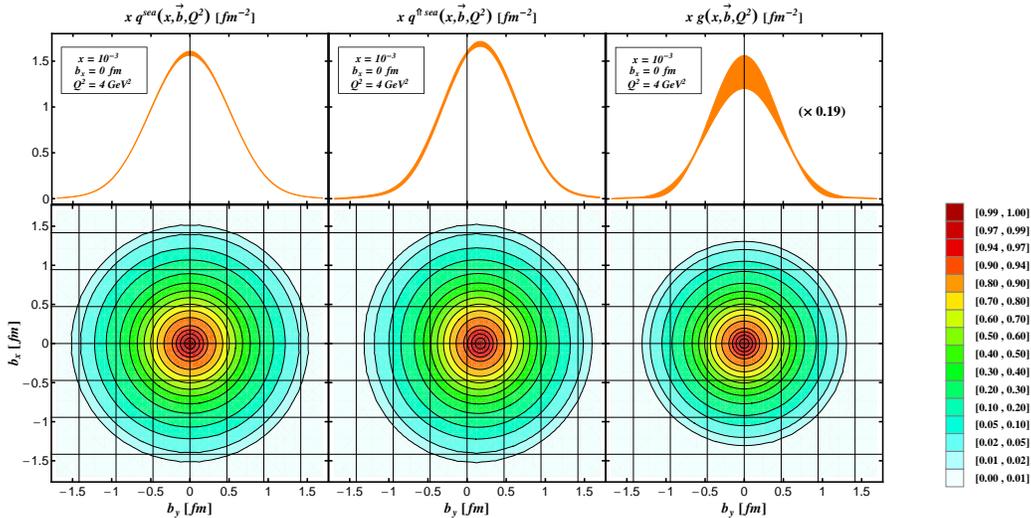}
\caption{Parton densities at $x=0.001$ and $\Q^2=4 \GeV^2$ versus impact parameter
$b$ were obtained from a combined least-squares fit to the HERA collider and EIC pseudo
data in LO approximation: relative densities (lower row) and their values at $b_x=0$ for the unpolarized sea
quark parton densities of a unpolarized proton (left),
a transversely polarized proton (middle), and the unpolarized gluon parton density of
a unpolarized proton (right), its value is rescaled by a factor 0.19. Figure is taken from \cite{Aschenauer:2013qpa}.
}
\label{fig:EIC}       
\end{figure}

New experiments with dedicated detectors are planned at COMPASS II and JLAB@12$\GeV$, which will fill the kinematical gap between HERMES and H1/ZEUS experiments  as well as between previous JLAB@6$\GeV$ and HERMES experiments. In particular it is expected that the high-luminosity experiments at JLAB will provide a large and high precision data set which will be very important for a global GPD analysis. It is worth to mention that having a transverse polarized target at COMPASS would allow to address the sea quark and gluon content of the GPD $E$ and this provides a model dependent access to the angular momentum of these parton species. Proposed electron-proton/ion scattering experiments such as the Large-Hadron-Electron-Collider at CERN \cite{AbelleiraFernandez:2012cc} and a Electron-Ion-Collider (EIC) \cite{Deshpande:2012bu} would allow to measure exclusive processes in the very small-$\xB$ region and in the small-$\xB$ region for polarized target. In particular the potential of DVCS measurements on an EIC has been studied in more detail \cite{Aschenauer:2013qpa} and revealed that one can access besides the GPD $H$ also the GPD $E$. In Fig.~\ref{fig:EIC} the model dependent imagining is sketched, where it was assumed that $t$-dependence is the same for all SO(3)-PWs.
It should be also emphasized that besides time-like DVCS measurements at JLAB also J-PARC (Japan Proton Accelerator Research Complex) offers the possibility to study GPDs in time-like DVMP. As argued in \cite{Kumano:2009he}, also pure hadronic reaction might be interesting for phenomenology. Another interesting possibility is to access GPDs in hard exclusive processes that are measured in neutrino experiments, see discussions in \cite{Psaker:2006gj,Kopeliovich:2012dr,Kopeliovich:2013ae,Kopeliovich:2014pea,Kopeliovich:2014jva}, which allow for a flavor separation in weak DVCS
experiments. DVCS and DVMP experiments with a neutron target \cite{Mazouz:2007aa} provides another possibility to address the problem of flavor separation. These deeply virtual processes can be also measured with a nuclei target, which has its own interest.

Finally, let us emphasize that the GPD analysis of present and future measurements is much more intricate as for inclusive processes and that we already witnessed over interpretations of measurements, driven by the wish to address the two main goals of GPD phenomenology. Surely, to provide a GPD interpretation of high precision data, expected in the future, one should  not only be able to describe data rather one needs models that can be considered as realistic and respect also  positivity constraints. This problem can be only overcome in the wave function overlap representation, where {\em duality} can be used to obtain the complete GPD  from a diagonal parton overlap. Although
some work has been done in this direction \cite{Hwang:2007tb,Mueller:2011bk,Hwang:2012qua}, such models are at present not flexible enough for fitting.  In principle, such a field theoretical framework allows also to include transverse momentum dependence and points in the direction of a unifying description of hard inclusive, semi-inclusive, and exclusive processes \cite{Hwang2014}.

\begin{acknowledgements}
I am indebted to M.~Polyakov, A.~Radyushkin,  K.~Semenov-Tian-Shansky, O.~Teryaev, and Ch.~Weiss for discussions on GPD representations,
which stimulated me to summarize what has been known and to apply it for specific DD representations. I like to thank M.~Guidal and P.~Kroll
for discussions on the VGG and GK models, respectively.
\end{acknowledgements}



\end{document}